\documentclass[leqno,11pt,letter]{amsart}

\usepackage[full]{textcomp}
\usepackage{newpxtext}
\usepackage{times}
\usepackage{cabin} 
\usepackage{amsmath,amssymb} 
\usepackage[varqu,varl]{inconsolata} 
\usepackage[bigdelims,vvarbb]{newpxmath} 
\usepackage[cal=cm,bb=esstix,bbscaled=1.05]{mathalfa} 
\usepackage[bottom=.82in,top=.85in,left=1.15in,right=1.15in]{geometry}
\usepackage{savesym}
\let\savedegree\bigtimes
\let\bigtimes\relax
\usepackage{mathabx}
\let\bigtimes\savedegree
\usepackage[usenames,dvipsnames]{color}
\usepackage{hyperref}
\hypersetup{
 colorlinks,
 linkcolor={blue!90!black},
 citecolor={red!80!black},
 urlcolor={blue!50!black}
}
\usepackage{tikz}
\usetikzlibrary{decorations.markings}
\usepackage{mathrsfs}
\usepackage[all,cmtip]{xy}
\usepackage{mleftright}
\usepackage{booktabs}
\usepackage[foot]{amsaddr}

\usepackage{cite}
\usepackage{enumitem}

\usepackage{graphicx}
\usepackage{subfigure}
\usepackage{caption}
\usepackage{placeins}
\usepackage{comment}
\usepackage[normalem]{ulem}

\setlist[enumerate]{labelsep=*, leftmargin=1.5pc}
\setlist[enumerate]{label=\normalfont(\roman*), ref=\roman*}

\newtheorem{theorem}{Theorem}[section]

\theoremstyle{definition}

\newtheorem{remark}[theorem]{Remark}
\newtheorem{definition}[theorem]{Definition}

\numberwithin{equation}{section}

\newcommand{\IGNORE}[1]{}
\newcommand{\ignore}[1]{}
\newcommand{\Hilbert}{H}
\newcommand{\Projection}{P}

\newcommand{\la}{\langle}
\newcommand{\ra}{\rangle}
\newcommand{\sgn}{\operatorname{sgn}}
\newcommand{\bds}{\boldsymbol}

\newcommand{\ph}{\text{phys}}

\newcommand{\mf}{\mathfrak}
\newcommand{\td}{\tilde}
\newcommand{\csch}{\operatorname{csch}}
\newcommand{\sech}{\operatorname{sech}}
\newcommand{\sfc}{\text{s}}
\newcommand{\btm}{\text{b}}
\newcommand{\qua}{\text{qua}}
\newcommand{\per}{\text{per}}
\newcommand{\mfq}{\mathfrak{q}}

\newcommand{\veps}{\epsilon}

\newcommand{\re}{\operatorname{Re}}
\newcommand{\im}{\operatorname{Im}}

\newcommand{\mbb}[1]{\mathbb{#1}}

\newcommand{\mc}[1]{\mathcal{#1}}

\newcommand{\jd}{\displaystyle}

\newcommand{\der}[2]{\frac{\partial #1}{\partial #2}}

\newcommand{\e}[1]{{(#1)}}
\newcommand{\option}{\operatorname{op}}

\usepackage{xcolor}
\usepackage{etoolbox}
\makeatletter
\pretocmd\@bibitem{\color{black}\csname keycolor#1\endcsname}{}{\fail}
\newcommand\citecolor[1]{\@namedef{keycolor#1}{\color{blue}}}
\makeatother

\begin{document}

\author[Jon Wilkening and Xinyu Zhao]{Jon Wilkening and Xinyu Zhao}
\address{Department of Mathematics, University of California, Berkeley,
  Berkeley, CA 94720, USA}
\email{wilkening@berkeley.edu}
\address{Department of Mathematics and Statistics, McMaster University,
  Hamilton, Ontario, Canada L8S 4K1}
\email{zhaox171@mcmaster.ca}

\subjclass[]{}
\title{Spatially quasi-periodic water waves of finite depth}

\begin{abstract}
 We present a numerical study of spatially quasi-periodic gravity-capillary waves
of finite depth in both the initial value problem and traveling wave
settings. We adopt a quasi-periodic conformal mapping
formulation of the Euler equations, where one-dimensional
quasi-periodic functions are represented by periodic functions on a
higher-dimensional torus. We compute the time evolution of free
surface waves in the presence of a background flow and a
quasi-periodic bottom boundary and observe the formation of
quasi-periodic patterns on the free surface. Two types of
quasi-periodic traveling waves are computed: small-amplitude waves
bifurcating from the zero-amplitude solution and larger-amplitude waves
bifurcating from finite-amplitude periodic traveling waves. We derive
weakly nonlinear approximations of the first type and investigate the
associated small-divisor problem. We find that waves of the second
type exhibit striking nonlinear behavior, e.g., the peaks
and troughs are shifted non-periodically from the corresponding
periodic waves due to the activation of quasi-periodic modes.
\end{abstract}

\maketitle
\markboth{J. WILKENING AND X. ZHAO}{QUASI-PERIODIC WATER WAVES OF FINITE DEPTH}


\section{Introduction}
\label{sec:intro}

Free surface waves on incompressible fluids arise in many contexts in
fluid dynamics. Examples include ocean wave forecasting
\cite{janssen2003nonlinear, toffoli2008surface}, modeling the motion
of flows over obstacles and varying bottom
boundaries~\cite{viotti2014conformal, flamarion2020time,
  ambrose2022numerical}, and studying wind-wave interactions in
extreme wave events, such as freak waves
\cite{kharif2008windwave}. These models are described by the Euler
equations, which are usually studied under periodic boundary
conditions or the assumption that solutions decay to zero at
infinity~\cite{johnson97, alazard2014cauchy,
  harrop2017finite}. However, these assumptions are insufficient in
many problems of interest. For instance, a periodic wave could
interact with a bottom boundary with a different spatial  
period, or
subharmonic perturbations of a periodic traveling wave can grow in
amplitude, leading to quasi-periodic waves. To tackle these issues,
we recently proposed methods \cite{quasi:trav, quasi:ivp} to study the
Euler equations under quasi-periodic boundary conditions;
specifically, we studied spatially quasi-periodic waves of infinite
depth in two dimensions and developed numerical algorithms to compute
such waves. In this paper, we extend this previous work to the
finite-depth case and discuss both the initial value and traveling
wave problems in the quasi-periodic setting.

Finite-depth water waves exhibit interesting nonlinear dynamics.
It has been shown numerically that Fermi-Pasta-Ulam recurrence can occur
 in free surface waves of finite depth when
the wave amplitude is less than about $1/10$ of the fluid depth \cite{ruban2011numerical, ruban2012fermi}. 
A varying bottom boundary can lead to substantial amplifications of water waves.
There have been both experimental and numerical studies demonstrating increased 
freak wave activities when waves propagate over a sloping bottom, from a deeper to a shallower domain \cite{trulsen2012laboratory, ducrozet2017influence}.
In the problem of long waves approaching vertical walls, an abrupt transition in the bottom boundary can cause large runups on the wall or
wave breaking, which generally occurs when the wave crest overturns \cite{viotti2014extreme, herterich2019extreme}.
The interaction between a rotational wave current and a varying bottom boundary gives rise to 
a time-dependent Kelvin cat-eye structure \cite{flamarion2020time}.

The quasi-periodic dynamics of water waves have recently drawn
considerable attention.  Berti and Montalto \cite{berti2016quasi},
Baldi \emph{et al.} \cite{baldi2018time}, Berti \emph{et al.}
\cite{berti2021traveling, berti2021pure}, and Feola and Giuliani
\cite{feola2020trav} have used Nash-Moser theory to prove the
existence of small-amplitude temporally quasi-periodic water waves. On
the numerical side, Wilkening computed new families of
relative-periodic \cite{collision} and traveling-standing
\cite{waterTS} water waves.  Although the physical mechanisms
  are different, temporally and spatially quasi-periodic waves have
  similar mathematical structures in that they can both be formulated
  in terms of periodic functions on a higher-dimensional torus.
Damanik and Goldstein \cite{damanik2016kdv}
proved the global existence and uniqueness of small-amplitude
spatially quasi-periodic solutions of the KdV equation.  Oh
\cite{oh2015nonlinear} and Dodson \emph{et al.}
\cite{dodson2020nonlinear} showed the local existence of spatially
quasi-periodic solutions of nonlinear Schr{\"o}dinger equations.

We were originally motivated by the structure of quasi-crystals in
material science.  Blinov \cite{blinov2015periodic} used
quasi-periodic solutions of the Schr\"{o}dinger equation to describe
the electronic structure of non-interacting electrons in a
quasi-crystals.  To study how electrons move through quasi-crystals,
Torres \emph{et al.} \cite{torres2003quasiperiodic} created
quasi-periodic standing waves by vibrating a fluid-filled pan with a
quasi-periodic bottom boundary and sent a transverse wave pulse across
the fluid that develops a non-periodic pattern in which the
spacing between the wave peaks is not constant.  Their observation
inspired us to ask the following question: how do we compute the exact
dynamics of free surface waves in the presence of a quasi-periodic
bottom boundary?  To address this question, one needs to study the
free surface wave problem in a quasi-periodic framework.

Another reason for our interest in quasi-periodic water waves originates from the dispersion relation of gravity-capillary waves of finite depth:
\begin{equation}\label{eq:dispersion_relation}
c^2 = (gk^{-1} + \tau k)\tanh(k h).
\end{equation}
Here $c$ is the phase speed, $k$ is the wave number, $g$ is the
acceleration due to gravity, $\tau$ is the coefficient of surface
tension and $h$ is the depth of the fluid. It is known
\cite{trichtchenko:16} that when $\tau/(gh^2) < 1/3$, there exists
$c_{\text{crit}}$ between 0 and $\sqrt{gh}$ such that for any fixed
phase speed $c>c_\text{crit}$, there are two distinct positive wave
numbers satisfying the dispersion
relation~(\ref{eq:dispersion_relation}), which we denote by $k_1$ and
$k_2$.  Any superposition of waves with these two wave numbers is a
solution of the linearized traveling wave problem.  If $k_1$ and $k_2$
are rationally related, the linear solution is spatially periodic and
related to the well-studied Wilton ripples \cite{wilton1915,
  trichtchenko:16, akers2019periodic, akers2021wilton}.  On the other
hand, if $k_1$ and $k_2$ are irrationally related, the linear solution
will be spatially quasi-periodic, which gives a natural place to
search for nonlinear quasi-periodic traveling solutions.  Bridges and
Dias \cite{bridges1996spatially} first studied these spatially
quasi-periodic traveling waves using a spatial Hamiltonian structure
and constructed weakly nonlinear approximations of these waves.
Recently, we \cite{quasi:trav} used a conformal mapping formulation of
the water wave equations and computed highly accurate numerical
solutions of the fully nonlinear problem in the case of deep water.
These computations are performed on a two-dimensional torus from
  which we extract 1D quasi-periodic functions via $u(\alpha) = \tilde
  u(k_1\alpha, k_2\alpha)$.  The computational challenges are similar
  to those of computing time-periodic standing waves
  \cite{mercer:92,water1,wilkening2012overdetermined}. The main
  difference is that standing waves can be formulated as a nonlinear
  two-point boundary value problem, which reduces the number of
  unknowns from $O(N^2)$ to $O(N)$ initial degrees of freedom, while
  quasi-periodic traveling waves have $O(N^2)$ unknowns but a simpler
  objective function whose main cost is the relatively inexpensive
  two-dimensional FFT.
 In the present work, we aim to further extend these techniques to the
case of finite-depth water.

Following \cite{quasi:trav, quasi:ivp}, we adopt a conformal mapping formulation of the free surface Euler equations \cite{dyachenko1996analytical, dyachenko1996nonlinear, choi1999exact, dyachenko2001dynamics, zakharov2002new, li2004numerical, hunter2016two, dyachenko:S:2019}.
In the finite depth case, the fluid domain with a curved surface and an uneven bottom boundary is mapped conformally onto a horizontal strip instead of the lower half-plane. Since the conformal mapping depends on time, even though the bottom boundary is fixed in physical space, 
the representation of the bottom boundary in conformal space varies with time.  
Ruban \cite{ruban2004water, ruban2005water} fixed the width of the strip and used a composition
of two conformal mappings to map the strip to the fluid domain -- the first leaves the real axis invariant and the second maps the real line to the bottom boundary.  
Viotti \emph{et~al.} \cite{viotti2014conformal}, Flamarion \emph{et al.} 
\cite{flamarion2021iterative, Flamarion:2019:Rotational} and Ribeiro \emph{et al.}
\cite{Ribeiro:2017:Rotational} let the width of the strip vary with time to keep the wave length the same in physical space and conformal space. 
They used a fixed-point iterative method to compute the bottom profile at different times. 
In order that water waves possess the same quasi-periods in both physical and conformal spaces, we also let the strip width be a time-dependent variable. However, in contrast to \cite{viotti2014conformal, flamarion2021iterative}, we 
compute the time evolution of the bottom profile directly, employing analytical properties of the conformal mapping, similar to \cite{ruban2004water, ruban2005water}.  
As in the infinite-depth case \cite{quasi:ivp, quasi:trav}, we introduce finite-depth quasi-periodic Hilbert transforms to 
relate the real and imaginary parts of the conformal mapping and to compute the kinematic boundary condition on the free surface. 
These Hilbert transforms are Fourier multiplier operators and are easier to compute in a quasi-periodic setting than a more direct computation of the Dirichlet-Neumann operator \cite{craig:sulem} in physical space, e.g., using boundary integral methods \cite{ambrose2022numerical}.

In computing the dynamics of free surface waves over a varying bottom boundary, it is usually assumed that the spatial periods of the free surface wave and the bottom boundary are
the same or one is an integer multiple of the other.
In this paper, we study a new situation where their spatial periods are irrationally related.  
Specifically, in one of the examples presented in Section \ref{sec:ivp}, we compute the time evolution of an initially periodic free surface wave with period $2\pi$ in the presence of a periodic bottom boundary with period $2\sqrt{2} \pi$.
We find that the periodic wave becomes a quasi-periodic wave, with each wave peak and trough evolving differently as it interacts with the bottom boundary. 
We also compute the time evolution of an initially flat free surface in the presence of a background flow and a quasi-periodic bottom boundary.  Similar to the experiment by Torres \emph{et al.} \cite{torres2003quasiperiodic}, we also observe that the free surface wave develops quasi-periodic patterns as a result of interactions between the background flow and the quasi-periodic bottom boundary. The wave peaks and troughs are asymmetric and the distance between adjacent wave peaks is not constant. 

In Section \ref{sec:num:trav}, we compute two types of  quasi-periodic traveling solutions: waves that bifurcate from the zero-amplitude solution and waves that bifurcate from finite-amplitude periodic traveling solutions. 
For the first type, we use linearization about the zero solution for the initial bifurcation direction and obtain a three-parameter family of solutions prescribed by the fluid depth and Fourier coefficients corresponding to wave numbers $k_1$ and $k_2$; these are called the base Fourier coefficients.
Similar to the case of deep water \cite{quasi:trav}, when the amplitudes of the base Fourier coefficients are small, the solutions are of small amplitude and are close to the linear solution. For the second type, we linearize the governing equations around a finite-amplitude $2\pi$-periodic traveling wave. For the bifurcation direction in this case, we use a quasi-periodic function of the following form in the kernel of the linear operator:
\begin{equation}
\delta \eta(\alpha) = e^{ik\alpha} \eta_0(\alpha) + c.c.,
\end{equation}
where $\eta_0$ possesses the same wavelength as the periodic traveling wave, the notation $c.c.$ denotes the complex conjugate of the preceding term, and we set $k = 1/\sqrt{2}$ in this paper. 
This method has also been used to compute secondary periodic bifurcations with $k = 1/2$ and $k = 1/3$ by 
Chen and Saffman \cite{chen1980numerical} and with $k = 1/9$ by Vanden-Broeck \cite{vanden2014periodic}.
In the present work, we obtain quasi-periodic traveling waves that bifurcate from a periodic traveling wave whose first Fourier mode resonates with the fifth Fourier mode.
The periodic traveling wave is a solution of the Wilton ripple problem and the wave peaks look like ``cat ears''. The bifurcated wave still preserves this characteristic; however, influenced by the Fourier modes in the quasi-periodic direction, the distance between the successive ``ears'' is no longer constant. 

The paper is organized as follows.  In
Section~\ref{sec:equation:motion}, we define finite-depth
quasi-periodic Hilbert transforms and derive equations of motion for
quasi-periodic free surface waves in conformal space when the bottom
boundary is not necessarily flat. In Section~\ref{sec:traveling}, we
obtain the governing equations of quasi-periodic traveling waves in
the case of finite-depth water with a flat bottom boundary and
establish weakly nonlinear approximations of these waves and the role
of small divisors in computing successive approximations.  In Section
~\ref{sec:numerical}, we use a Fourier pseudo-spectral method to
compute solutions of the initial value and traveling wave problems and
present various numerical examples.  Following the idea in
\cite{quasi:ivp, quasi:trav}, we lift the one-dimensional
quasi-periodic problem to a higher-dimensional periodic torus where
the computation is performed.  We formulate the traveling wave problem
as an overdetermined nonlinear least-squares problem that we solve
through a variant of the Levenberg-Marquardt method
\cite{nocedal,wilkening2012overdetermined}. For the initial value
problem, we consider the natural setting where the quasi-periodic
initial condition and bottom boundary are posed in physical space.
We present a method of transforming them to conformal space in
Appendix \ref{sec:conformal:transform}.


\section{Equations of motion}\label{sec:equation:motion}



\subsection{Governing equations in physical space} \label{sec:govern:phys}

Gravity-capillary waves of finite depth are governed by the
free-surface Euler equations \cite{zakharov1968stability,johnson97}.
In two dimensions, they may be written as
\begin{equation} \label{eq:general:initial}
  \eta^\sfc(x, 0) = \eta^\sfc_0(x), \qquad \varphi(x, 0) = \varphi_0(x), \qquad t=0, \quad
  x\in \mathbb{R},
\end{equation}
\begin{equation}\label{eq:harmonic:function}
  \begin{aligned}
    \Phi_{xx} + \Phi_{yy} &= 0, &\qquad\qquad \eta^\btm(x) &< y < \eta^\sfc(x, t),\\
    \Phi &= \varphi, &\qquad\qquad y &= \eta^\sfc(x, t),\\
     \nabla \Phi \cdot \bds n &= 0, &\qquad\qquad y &= \eta^\btm(x),\\
  \end{aligned}
\end{equation}
\begin{equation}\label{eq:eta:evolve}
  \eta^\sfc_t = \Phi_y - \eta^\sfc_x\Phi_x, \qquad\qquad y = \eta^\sfc(x, t),
\end{equation}
\begin{equation}\label{eq:Bernoulli}
  \varphi_t = \Phi_y\eta^\sfc_t-\frac{1}{2}\Phi_x^2-\frac{1}{2}\Phi_y^2-
  g\eta^\sfc+\tau\frac{\eta_{xx}^\sfc}{\big(1+(\eta_x^\sfc)^2\big)^{3/2}}+C(t), \qquad y = \eta^\sfc(x, t),
\end{equation}
where $x$ is the horizontal coordinate, $y$ is the vertical
coordinate, $t$ is the time, $\Phi(x, y, t)$ is the velocity
potential of the fluid, $\eta^\sfc(x, t)$ is the free surface
elevation, $\eta^\btm(x)$ is the fixed bottom profile,
$g$ is the vertical acceleration due to gravity, and $\tau$ is the
coefficient of surface tension, which is zero for gravity waves.
 Equation (\ref{eq:eta:evolve}) is the kinematic boundary condition and 
(\ref{eq:Bernoulli}) is the dynamic boundary condition.
The function $C(t)$ in 
(\ref{eq:Bernoulli}) is an arbitrary integration constant that is allowed
to depend on time but not space. 
We are interested in the dynamics of the water waves in the presence of a varying bottom boundary;
in other words, the bottom profile is not a constant function. 
When the bottom boundary is flat, it is usually assumed that there is no background flow.
Indeed, in this case, the system is Galilean invariant, which means any background flow can
be eliminated by viewing the system in a moving frame.
However, this is not true when the bottom boundary is variable; the interaction between the background flow and the bottom boundary can lead to interesting nonlinear dynamics. 
Therefore, it is 
meaningful to incorporate a background flow in the problem description by including a secular growth term in the
velocity potential, which is otherwise spatially periodic or quasi-periodic.


\subsection{Finite-depth quasi-periodic Hilbert transforms}

As defined in \cite{moser1966theory,dynnikov2005topology},
a quasi-periodic, real-analytic function $f(\alpha)$ is a function
of the form
\begin{equation}\label{eq:general_quasi_form}
  f(\alpha) = \tilde f(\bds{k} \alpha), \qquad
  \tilde f(\bds\alpha) = \sum_{\boldsymbol{j}\in\mathbb{Z}^d}\hat{f}_{\boldsymbol{j}}
  e^{i\la\boldsymbol{j},\,\boldsymbol{\alpha}\ra}, \qquad
  \alpha\in\mbb R, \;\; \bds\alpha,\bds k \in \mathbb{R}^d,
\end{equation}
where $\la\cdot , \,\cdot\ra$ denotes the standard inner
product on $\mathbb{R}^d$ and $\tilde f$ is a periodic, real-analytic
function defined on the $d$-dimensional torus
 $\mathbb{T}^d := \mbb R^d\big/(2\pi\mbb Z)^d.$
We assume that $d \geq 2$ so that $f$ can be genuinely quasi-periodic.
Entries of the vector $\boldsymbol{k}$ are called the basic wave
numbers (or basic frequencies) of $f$ and are required to be linearly
independent over $\mathbb{Z}$.  Given a quasi-periodic function $f$,
the corresponding $\td f$ and $\bds k$ in
(\ref{eq:general_quasi_form}) are not unique. Indeed, if $\bds K$ is
any $d$-by-$d$ unimodular matrix, then $\tilde f'(\bds \alpha) =
\tilde f(\bds K \bds\alpha)$ also satisfies (\ref{eq:general_quasi_form})
with $\bds k' = \bds K^{-1} \bds k$.  For simplicity, we assume $\bds
k$ is given, along with $f$ or $\td f$, to pin down the
representation. Given $\bds k$, one can reconstruct $\td f$
and its Fourier coefficients $\hat f_{\bds j}$ from $f$ via
\begin{equation}\label{eq:fhat:from:f}
  \hat{f}_{\bds{j}} = \lim_{a\to\infty} \frac{1}{2a} 
  \int_{-a}^a f(\alpha) e^{-i\la \bds{j}, \bds{k}\ra\alpha} d\alpha,
  \qquad \bds{j}\in\mathbb{Z}^d.
\end{equation}
We refer to \cite{bohr:book} for detailed discussions of the above averaging formula.
We assume that $\tilde f(\bds\alpha)$ is real-analytic, which is
  equivalent to the conditions that $\hat f_{-\bds j}=\overline{\hat
    f_{\bds j}}$ for $\bds j\in\mbb Z^d$ and there exist positive
  numbers $M$ and $\gamma$ such that $|\hat f_{\bds j}|\le
  Me^{-\gamma\|\bds j\|}$, i.e., the Fourier modes $\hat{f}_{\bds{j}}$
  decay exponentially as $\|\bds j\|\rightarrow\infty$. 
  Next we introduce some operators that act on $f$ and $\td f$.

\begin{definition}
The projection operators $P$ and $P_0$ are defined by
\begin{equation}\label{eq:proj}
  \Projection = \operatorname{id} - \Projection_0, \qquad
  \Projection_0 [f] = \Projection_0 [\tilde f] = \hat{f}_{\boldsymbol{0}}
  = \frac{1}{(2\pi)^d} \int_{\mathbb{T}^d}
  \tilde f(\boldsymbol{\alpha})\, d\alpha_1\cdots\, d\alpha_d.
\end{equation}
Note that $\Projection$ projects onto the space of zero-mean functions
and $\Projection_0$ returns the mean value. There
  are two versions of $P$ and $P_0$, one acting on quasi-periodic
  functions defined on $\mbb R$ and one acting on torus functions
  defined on $\mbb T^d$.
 \end{definition}

\begin{definition}\label{def:deriv}
The derivative operator $\partial_\alpha$ that acts on $f$ or $\td f$
is defined by
\begin{equation}\label{eq:derivative}
 \partial_\alpha f(\alpha)= \partial_\alpha \td f(\bds k \alpha), \qquad 
 \partial_\alpha \td f(\bds \alpha) = 
 \sum_{\bds j \neq \bds 0}
i \la\bds j, \bds k \ra \hat f_{\bds j} e^{i \la\bds j, \bds \alpha \ra }.
\end{equation}
For simplicity of notation, we denote $\partial_\alpha f$ (or
  $\partial_\alpha \td f$) by $f_\alpha$ (or $\td f_\alpha$).  One can
also interpret $\partial_\alpha \td f$ as the directional derivative
of $\td f$ along the characteristic direction $\bds k$.
 \end{definition}

\begin{definition}\label{def:hilbert}
We introduce four finite-depth quasi-periodic Hilbert transforms
 $H^{\tanh}$, $H^{\coth}$, $H^{\csch}$, $H^{\sech}$
that act on $f$ and $\td f$ as follows  \cite{plotnikov01, viotti2014conformal}
\begin{equation}\label{eq:H:defs}
H^{\option}[f](\alpha) = H^{\option}[\td f](\bds k\alpha), \qquad
H^{\option}[\td f](\bds \alpha) = \sum_{\bds j \in\mbb Z^d}
i \hat{H}^{\option} \hat f_{\bds j} e^{i\la\bds j,  \bds\alpha\ra }, 
\end{equation}
where $\option = \tanh, \coth, \sech$ or $\csch$ and the symbol $\hat{H}^{\option}$
is given by
\begin{equation}\label{eq:H:symbol}
\begin{aligned}
\hat H_{\bds j}^{\tanh} &= i \tanh\big(\la\bds j, \bds k \ra h\big),& \qquad
\hat H_{\bds j}^{\coth} &= \begin{cases} (-i) \coth\big(\la\bds j, \bds k \ra h\big), & \bds j\ne \bds 0, \\ 0 & \bds j=\bds 0, \end{cases} \\
\hat H_{\bds j}^{\sech} &= \sech\big(\la\bds j, \bds k \ra h\big),& \qquad
\hat H_{\bds j}^{\csch} &= \begin{cases} i \csch\big(\la\bds j, \bds k \ra h\big) & \bds j\ne \bds 0, \\ 0 & \bds j=\bds 0. \end{cases}
\end{aligned}
\end{equation}
Here $h$ is a positive parameter that will be discussed in Section \ref{sec:conf:map}.

We notice that
\begin{equation}
\lim_{h\to \infty} \hat H_{\bds j}^{\tanh} = i\sgn\big( \la\bds j, \bds k \ra \big), \qquad \qquad 
\lim_{h\to \infty} \hat H_{\bds j}^{\coth} = -i\sgn\big( \la\bds j, \bds k \ra \big).
\end{equation}
The latter coincides with the quasi-periodic Hilbert transform
introduced in \cite{quasi:ivp, quasi:trav} in the case of deep water
while the former is its pseudo-inverse.
\end{definition}


\subsection{The quasi-periodic conformal mapping}
\label{sec:conf:map}

%
\begin{figure}[t]
\begin{centering}
\includegraphics[width=0.6\textwidth]{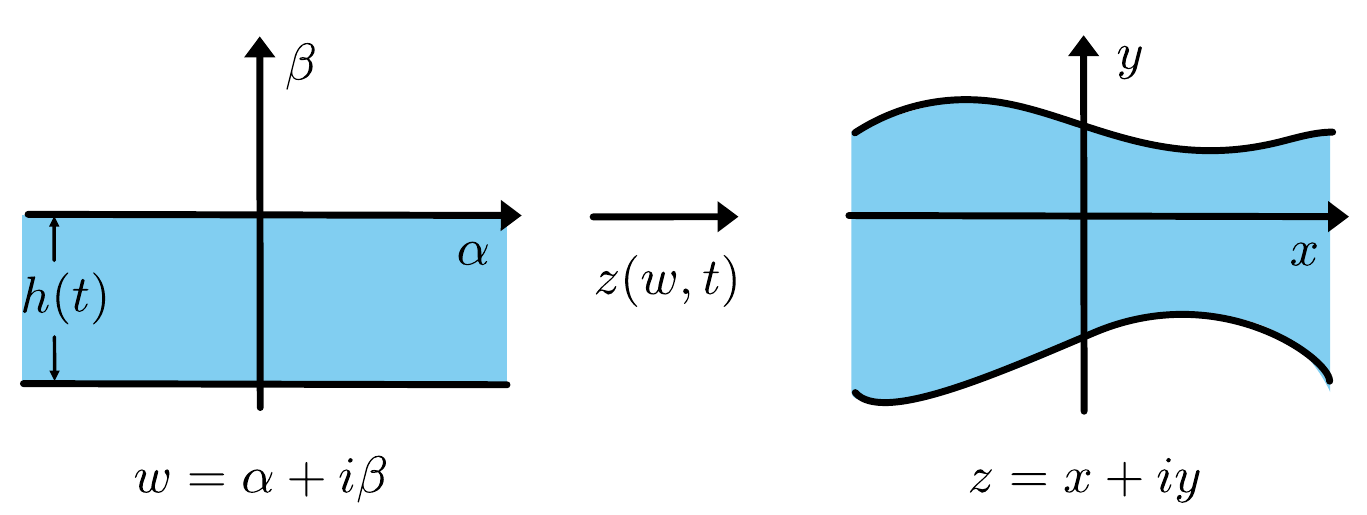}
\caption{\label{fig:conformal} 
The time-dependent conformal mapping.}
\end{centering}
\end{figure}

Figure \ref{fig:conformal} illustrates a time-dependent conformal mapping
\begin{equation} \label{eq:conformal_mapping}
  z(w, t) = x(\alpha, \beta, t) + i y(\alpha, \beta, t), \qquad\qquad w=\alpha+i\beta,
\end{equation}
that maps the infinite strip in the complex plane
\begin{equation}
S_{ h} = \{\alpha + i\beta\, : \, \alpha \in \mbb R, \,- h(t) < \beta <0\}
\end{equation}
to the fluid domain
\begin{equation}
\Omega_{f} = \{(x, y)\,: \, x \in\mbb R, \,  \eta^{\btm, \ph}(x) < y < \eta^{\sfc, \ph}(x, t)\}.
\end{equation}
To avoid ambiguity,  we use $\eta^{\sfc, \ph}$ and $\eta^{\btm, \ph}$ to denote the free surface elevation and the bottom profile in physical space, respectively, whereas $\eta^\sfc$ and $\eta^\btm$ are used as conformal variables henceforth. 

We assume that $z(w, t)$ can be extended continuously to $\overline {S_{ h}}$ and 
maps the top and bottom boundary of the strip to the free surface and the bottom boundary
 of the fluid domain, respectively. 
Denoting
\begin{equation}\label{eq:z:sfc:btm}
\begin{aligned}
\zeta^\sfc(\alpha, t) &=  z|_{\beta = 0}(\alpha, t) =  x(\alpha, 0, t) + i y(\alpha, 0, t) =
\xi^\sfc(\alpha, t) + i \eta^\sfc(\alpha, t), \\
\zeta^\btm(\alpha, t)&= z|_{\beta= - h(t)}(\alpha, t) =  x(\alpha, - h(t), t) + iy(\alpha, - h(t), t) =
\xi^\btm(\alpha, t) + i \eta^\btm(\alpha, t),
\end{aligned}
\end{equation}
we have
\begin{equation}\label{eq:eta:b:conf}
\eta^{\sfc, \ph}(\xi^\sfc(\alpha, t), t) = \eta^\sfc(\alpha, t), \qquad \qquad
\eta^{\btm, \ph}(\xi^\btm(\alpha, t)) = \eta^\btm(\alpha, t).
\end{equation}
For later use in the derivation of the governing equations in conformal space, 
we compute the derivative with respect to $\alpha$ and $t$ on both sides of 
(\ref{eq:z:sfc:btm}) and obtain that
\begin{equation}\label{eq:x:y:deriv:sfc}
\begin{aligned}
x_\alpha  = \xi^\sfc_\alpha, \qquad y_\alpha= \eta^\sfc_\alpha,  \qquad
x_t = \xi^\sfc_t, \qquad y_t = \eta^\sfc_t, \qquad (\beta = 0)
\end{aligned}
\end{equation}
as well as 
\begin{equation}\label{eq:x:y:deriv:btm}
\begin{aligned}
x_\alpha  = \xi^\btm_\alpha, \qquad y_\alpha= \eta^\btm_\alpha,  \qquad
y_\alpha h_t + x_t = \xi^\btm_t, \qquad  
-x_\alpha  h_t + y_t = \eta^\btm_t, \qquad 
(\beta = -h(t))
\end{aligned}
\end{equation}
where we use the Cauchy-Riemann relation $x_\alpha = y_\beta$ and $y_\alpha = -x_\beta$ in the last two equalities.
The derivative of (\ref{eq:eta:b:conf}) with respect to $\alpha$ and $t$ yields 
\begin{equation}\label{eq:deta:sfc}
\eta^{\sfc, \ph}_x \xi^\sfc_\alpha = \eta^\sfc_\alpha, \qquad \qquad 
\eta^{\sfc, \ph}_x \xi^\sfc_t  + \eta^{\sfc, \ph}_t = \eta^\sfc_t.
\end{equation}
and
\begin{equation}\label{eq:deta:btm}
\eta^{\btm, \ph}_x \xi^\btm_\alpha = \eta^\btm_\alpha, \qquad \qquad 
 \eta^{\btm, \ph}_x\xi^\btm_t =\eta^\btm_t.
\end{equation}

We are interested in the case where $\eta^\sfc$ and $\eta^\btm$ are quasi-periodic functions of the form (\ref{eq:general_quasi_form}),
\begin{equation}\label{eq:eta:sfc:btm:quasi}
\begin{aligned}
\eta^\sfc(\alpha, t) = \td \eta^\sfc(\bds k\alpha, t), \qquad
\td \eta^\sfc(\bds \alpha, t)= \sum_{\boldsymbol{j}\in\mathbb{Z}^d}\hat{\eta}_{\boldsymbol{j}}^\sfc(t) e^{i\la\bds j, \bds \alpha \ra}, \\
\eta^\btm(\alpha, t) = \td \eta^\btm(\bds k\alpha, t), \qquad
\td \eta^\btm(\bds \alpha, t)= \sum_{\boldsymbol{j}\in\mathbb{Z}^d}\hat{\eta}_{\boldsymbol{j}}^\btm(t) e^{i\la\bds j, \bds \alpha \ra},
\end{aligned}
\qquad \qquad \alpha\in\mbb R, \;\; \bds \alpha,\bds k \in \mathbb{R}^d,
\end{equation}
where $\bds k$ is fixed and its components are linearly independent
over $\mbb Z$.  This is different from the usual conformal mapping
framework \cite{meiron1981applications, dyachenko1996analytical,
  dyachenko2001dynamics, zakharov2002new, li2004numerical,
  milewski:10}, where $\eta^\sfc$ and, if present, $\eta^\btm$ are
assumed to be periodic.  Using the fact that $y$ is a harmonic
function defined on $S_{ h}$ and the boundary values of $y$ are given
by $y|_{\beta = 0} = \eta^\sfc$ and $y|_{\beta = - h} = \eta^\btm$, we
obtain that
\begin{equation}\label{eq:y:form}
y = \frac{1}{ h} \big(\hat \eta_{\bds 0}^\sfc - \hat \eta_{\bds 0}^\btm\big)\beta 
+ \hat \eta_{\bds 0}^\sfc
+ \sum_{\bds j\neq \bds 0} 
\frac{\sinh\big( \la\bds j, \bds k \ra(\beta+  h)\big)}{\sinh \big(\la\bds j, \bds k \ra h\big)}
\hat \eta_{\bds j}^\sfc e^{i\la\bds j, \bds k \ra \alpha}
-
\sum_{\bds j\neq \bds 0} 
\frac{\sinh\big( \la\bds j, \bds k \ra\beta\big)}{\sinh \big(\la\bds j, \bds k \ra h\big)}
\hat \eta_{\bds j}^\btm e^{i\la\bds j, \bds k \ra \alpha}.
\end{equation}
The harmonic conjugate of $y$, which is $x$, can be computed from (\ref{eq:y:form})
using the Cauchy-Riemann equations  $x_\alpha = y_\beta$, $x_\beta = -y_\alpha$,
\begin{equation}\label{eq:x:form}
x = \frac{1}{ h}\big( \hat{\eta}_{\bds 0}^\sfc  - \hat \eta_{\bds 0}^\btm\big)\alpha
+ x_0
- \sum_{\bds j\neq \bds 0} i
\frac{\cosh\big( \la\bds j, \bds k \ra(\beta+  h)\big)}{\sinh \big(\la\bds j, \bds k \ra h\big)}
\hat \eta_{\bds j}^\sfc  e^{i\la\bds j, \bds k \ra \alpha}
+
\sum_{\bds j\neq \bds 0} i 
\frac{\cosh\big( \la\bds j, \bds k \ra\beta\big)}{\sinh \big(\la\bds j, \bds k \ra h\big)}
\hat \eta_{\bds j}^\btm e^{i\la\bds j, \bds k \ra \alpha}.
\end{equation}
Here $x_0$ is an integration constant, depending on time only, that we
are free to choose. Given $\Omega_f$ at any time, to fix the mapping
$z$, we need to specify two free parameters: $h$ and $x_0$.  We set
\begin{equation}\label{eq:h:free:param}
  h = \hat \eta_{\bds 0}^\sfc - \hat\eta_{\bds 0}^\btm, \qquad\quad
  x_0 = 0.
\end{equation}
Hence, the first terms in (\ref{eq:y:form}) and (\ref{eq:x:form}) are
just $\alpha$ and $\beta$.  One can choose $ h$ in the same way when
the fluid domain is periodic in $x$ so that wavelengths do not change
under the conformal mapping \cite{viotti2014conformal}. Alternatively,
one may set $h=1$, as is done in \cite{ruban2004water,ruban2005water}
in the periodic case. Setting $x_0=0$ requires a certain choice to be
made for a parameter in the time evolution equations \cite{quasi:ivp};
this will be discussed in Section \ref{sec:govern:eq:conf}. Until then,
we leave $x_0(t)$ in the representation as a time-dependent parameter.

Comparing (\ref{eq:y:form}) and (\ref{eq:x:form}), we notice that the
values of $x$ and $y$ at the top and bottom boundary of $S_h$ are
related by the quasi-periodic Hilbert transforms of \eqref{eq:H:defs},
\begin{equation}\label{eq:xi:eta:hilb}
\begin{aligned}
\xi^\sfc(\alpha, t) &= \alpha + x_0(t) + H^{\coth}[\eta^\sfc](\alpha, t) 
+ H^{\csch}[\eta^\btm](\alpha, t), \\
\xi^\btm(\alpha, t) &= \alpha + x_0(t) - H^{\csch}[\eta^\sfc](\alpha, t) - H^{\coth}[\eta^\btm](\alpha, t).
\end{aligned}
\end{equation}
The corresponding torus functions are given in
  \eqref{eq:xi:tilde} below.


\subsection{The quasi-periodic complex velocity potential}

Let $\Phi^\ph(x,y,t)$ denote the velocity potential in physical space
from Section~\ref{sec:govern:phys} and let
$W^\ph(x+iy,t)= \Phi^\ph(x,y,t)+i\Psi^\ph(x,y,t)$ be the
complex velocity potential, where $\Psi^\ph$ is the stream function.
Using the conformal mapping (\ref{eq:conformal_mapping}), we pull
back these functions to the strip $S_{ h}$ and define
\begin{equation}
  W(w,t) = \Phi(\alpha,\beta,t)+i\Psi(\alpha,\beta,t) = W^\ph(z(w,t),t), \qquad\quad
  w=\alpha+i\beta.
\end{equation}
We denote $\varphi^\sfc = \Phi\vert_{\beta =0 }$, $\varphi^\btm = \Phi\vert_{\beta = - h}$,
$\psi^\sfc = \Psi\vert_{\beta =0 }$, $\psi^\btm = \Psi\vert_{\beta = - h}$
and use (\ref{eq:z:sfc:btm}) to obtain
\begin{equation}\label{eq:varphi:psi:bdr}
\begin{aligned}
\varphi^\sfc(\alpha, t)&= \Phi^\ph(\xi^\sfc(\alpha, t), \eta^\sfc(\alpha, t), t) 
= \varphi^{\sfc, \ph}(\xi^\sfc(\alpha, t), t), \\
\varphi^\btm(\alpha, t)&= \Phi^\ph(\xi^\btm(\alpha, t), \eta^\btm(\alpha, t), t) , \\
\psi^\sfc(\alpha, t)&= \Psi^\ph(\xi^\sfc(\alpha, t), \eta^\sfc(\alpha, t), t) 
= \psi^{\sfc, \ph}(\xi^\sfc(\alpha, t), t), \\
\psi^\btm(\alpha, t)&= \Psi^\ph(\xi^\btm(\alpha, t), \eta^\btm(\alpha, t), t) , 
\end{aligned}
\end{equation}
where $\varphi^{\sfc, \ph}$, $\psi^{\sfc, \ph}$ represent the values of $\Phi^\ph$ and $\Psi^\ph$ on the free surface.  
Following \cite{viotti2014conformal} for the periodic case, we assume that there is a background flow of horizontal mean velocity $\mc U$ and the quasi-periodic part of $\varphi^\sfc$ has the same quasi-periods as 
$\eta^\sfc$ and $\eta^\btm$ 
\begin{equation}\label{eq:varphi:sfc}
 \varphi^\sfc(\alpha, t) = \mc U\alpha + \tilde{\varphi}^\sfc(\bds{k}\alpha, t), \qquad
\tilde \varphi^\sfc(\bds \alpha, t) = \sum_{\boldsymbol{j}\in\mathbb{Z}^d}\hat{\varphi}_{\boldsymbol{j}}^\sfc(t) e^{i\la\bds j, \bds \alpha \ra},
  \qquad
 \alpha\in\mbb R, \;\; \bds \alpha,\bds k \in \mathbb{R}^d.
\end{equation}
According to (\ref{eq:harmonic:function}),
the bottom boundary is a streamline, therefore $\psi^\btm$ is a constant function (or a function of time only).  
Considering that adding constants (or
  functions of time) to $\Phi$ and $\Psi$ will not affect the fluid motion, we set $\hat \varphi^\sfc_{\bds 0} = 0$
  and 
\begin{equation}\label{eq:psi:btm}
 \psi^\btm = 0.
\end{equation} 
Since $\Phi$ and $\Psi$ are harmonic conjugates satisfying
boundary conditions (\ref{eq:varphi:sfc}) and (\ref{eq:psi:btm}), 
we obtain 
\begin{equation}\label{eq:Phi:Psi}
\begin{aligned}
\Phi &= \mc U\alpha + \sum_{\bds j \neq \bds 0} \hat\varphi_{\bds j}
\frac{\cosh\big(\la\bds j, \bds k \ra(\beta +  h)\big)}{\cosh\big(\la\bds j, \bds k \ra  h\big)}
e^{i\la\bds j, \bds k \ra \alpha}, \\
\Psi & = (\beta +  h)\mc U + \sum_{\bds j \neq \bds 0} i\hat\varphi_{\bds j}
\frac{\sinh\big(\la\bds j, \bds k \ra(\beta +  h)\big)}{\cosh\big(\la\bds j, \bds k \ra  h\big)}
e^{i\la\bds j, \bds k \ra \alpha}.
\end{aligned}
\end{equation}
Comparing the values of $\Phi$ and $\Psi$ at $\beta = 0$ and $\beta = \mf - h$, we conclude that 
\begin{equation}\label{eq:varphi:psi:hilb}
\begin{gathered}
\varphi^\sfc(\alpha, t) = \mc U\alpha + H^{\coth}[\psi^\sfc](\alpha, t), \qquad 
\psi^\sfc_\alpha(\alpha, t) = H^{\tanh}[\varphi^\sfc_\alpha](\alpha, t), \\
\varphi^\btm_\alpha(\alpha, t) =  \mc U + H^{\sech}[\varphi^\sfc_\alpha](\alpha, t) 
= \mc U - H^{\csch}[\psi^\sfc_\alpha](\alpha, t).
\end{gathered}
\end{equation}


\subsection{Governing equations in conformal space}
\label{sec:govern:eq:conf}

We now present a derivation of the equations of motion for quasi-periodic surface water waves
in a conformal mapping formulation when the fluid is of finite depth and the bottom boundary is not necessarily flat.
This is an extension of the results in \cite{quasi:ivp}, 
where the fluid depth is infinite.  
Since the conformal mapping is time-dependent, even though the bottom profile in physical space is fixed, the width of the strip in the conformal domain and the parameterization of the bottom boundary in conformal space, denoted $h(t)$ and $\zeta^\btm(\alpha, t)$, respectively, both vary with time. 
Therefore besides the free surface, the time evolution equations of $h$ and $\zeta^\btm$ in conformal space are needed to describe the evolution of the fluid domain. 
This is the main difference between the conformal mapping formulations in deep and finite-depth water.

To begin, we  use the chain rule to obtain
\begin{equation} \label{eq:potential:chain:rule}
  \frac{dW}{dw} = \frac{dW^\ph}{dz}\cdot\frac{dz}{dw} \qquad \Rightarrow \qquad
\Phi^\ph_x + i \Psi^\ph_x = \frac{\Phi_\alpha + i\Psi_\alpha}{x_\alpha+iy_\alpha}.
\end{equation}
Since $\Phi^\ph_y = -\Psi^\ph_x$, we can express the velocity of the fluid, which is the
gradient of $\Phi^\ph$,
in terms of $\Phi_\alpha$ and $\Psi_\alpha$
\begin{equation} \label{eq:velocity}
\Phi_x^\ph = \frac{\Phi_\alpha x_\alpha + \Psi_\alpha y_\alpha}{x_\alpha^2 + y_\alpha^2},
\qquad \qquad
\Phi_y^\ph = \frac{\Phi_\alpha y_\alpha - \Psi_\alpha x_\alpha}{x_\alpha^2 + y_\alpha^2}.
\end{equation}
Evaluating (\ref{eq:velocity}) on the free surface,
we have
\begin{equation} \label{eq:velocity:sfc}
\left.\Phi^\ph_x\right\vert_{z = \zeta^\sfc(\alpha, t)} 
  = \frac{\varphi_\alpha^\sfc \xi_\alpha^\sfc + \psi_\alpha^\sfc \eta_\alpha^\sfc}{J^\sfc}, \quad
\left.\Phi^\ph_y\right\vert_{z = \zeta^\sfc(\alpha, t)} 
 = \frac{\varphi_\alpha^\sfc \eta_\alpha^\sfc - \psi_\alpha^\sfc \xi_\alpha^\sfc}{J^\sfc},
 \quad   J^\sfc = (\xi_\alpha^\sfc)^2 + (\eta_\alpha^\sfc)^2.
\end{equation}

Next we derive the kinematic boundary condition in conformal space. 
We define the function 
\begin{equation}\label{eq:zt:by:zw}
\vartheta := \frac{z_t}{z_w} = \frac{x_tx_\alpha + y_t y_\alpha }{x_\alpha^2 + y_\alpha^2} + 
i \frac{y_t x_\alpha - x_ty_\alpha }{x_\alpha^2 + y_\alpha^2},
\end{equation}
which is holomorphic on $S_{ h}$ 
as long as $z_w$ is bounded away from zero. 
Evaluating (\ref{eq:zt:by:zw}) at  $\beta = 0$ and $\beta = -h(t)$ and replacing 
the derivatives of $x$ and $y$ by the derivatives of $\xi^\sfc$ and $\eta^\sfc$ using 
(\ref{eq:x:y:deriv:sfc}), (\ref{eq:x:y:deriv:btm}), we obtain that
\begin{equation}\label{eq:zt:by:zw:sfc}
\re \vartheta\Big\vert_{\beta = 0} 
= \frac{\xi_t^\sfc\xi_\alpha^\sfc +  \eta_t^\sfc\eta_\alpha^\sfc}{J^\sfc},
\qquad \qquad
\im \vartheta\Big\vert_{\beta = 0}
 = \frac{\eta_t^\sfc \xi_\alpha^\sfc  - \xi_t^\sfc \eta_\alpha^\sfc }{J^\sfc},
\end{equation}
\begin{equation}\label{eq:zt:by:zw:btm}
\begin{aligned}
\re \vartheta\Big|_{\beta = - h(t)} 
&= \frac{ \xi_t^\btm\xi_\alpha^\btm + \eta_t^\btm \eta_\alpha^\btm }{J^\btm}, \\
\im \vartheta\Big|_{\beta = - h(t)}
 &= \frac{\eta_t^\btm \xi_\alpha^\btm  - \xi_t^\btm \eta_\alpha^\btm }{J^\btm}  +  h_t,
 \end{aligned}
 \qquad \qquad
 J^\btm = (\xi_\alpha^\btm)^2 + (\eta_\alpha^\btm)^2.
\end{equation}
Furthermore, the substitution of (\ref{eq:deta:sfc}) and (\ref{eq:velocity:sfc}) 
into (\ref{eq:eta:evolve}) gives
\begin{equation}\label{eq:kinematic:sfc}
\eta^\sfc_t \xi^\sfc_\alpha - \xi^\sfc_t\eta^\sfc_\alpha= -\psi^\sfc_\alpha.
\end{equation}
Therefore we have
\begin{equation}\label{eq:zt:zw:im:sfc}
\im \vartheta \Big\vert_{\beta = 0} = -\frac{\psi_\alpha^\sfc}{J^\sfc}.
\end{equation}
Substituting (\ref{eq:deta:btm}) into (\ref{eq:zt:by:zw:btm}),
we obtain that $\eta_t^\btm \xi_\alpha^\btm  - \xi_t^\btm \eta_\alpha^\btm = 0$, thus
\begin{equation}\label{eq:zt:zw:im:btm}
\im \vartheta\Big|_{\beta = - h(t)}
 =  h_t.
 \end{equation}
Since $ h_t$ does not depend on the spatial variable,  similar to (\ref{eq:varphi:psi:hilb}), 
$\re \vartheta \vert_{\beta = 0}$ and $\re \vartheta\vert_{\beta = - h(t)}$
can be determined by $\im \vartheta\vert_{\beta = 0}$ up to an additive constant (that may depend on time but not space) as follows,
\begin{equation}\label{eq:kinematic:hilb}
\frac{\xi_t^\sfc\xi_\alpha^\sfc +  \eta_t^\sfc\eta_\alpha^\sfc}{J^\sfc} 
= -H^{\coth}\bigg[\frac{\psi_\alpha^\sfc}{J^\sfc}\bigg] + C_1, \qquad \qquad
\frac{ \xi_t^\btm\xi_\alpha^\btm + \eta_t^\btm \eta_\alpha^\btm }{J^\btm}
= H^{\csch}\bigg[\frac{\psi_\alpha^\sfc}{J^\sfc}\bigg] + C_1.
\end{equation}
Since $\vartheta$ is a holomorphic function defined on $S_h$, using Cauchy's integral theorem, we obtain
\begin{equation}\label{eq:cauchy:q}
\int_{-a+i(\epsilon-h)}^{a + i(\epsilon-h)} + \int_{a+i(\epsilon-h)}^{a - i\epsilon}
+ \int_{a-i\epsilon}^{-a - i\epsilon} + \int_{-a-i\epsilon}^{-a +i(\epsilon-h)}
\vartheta(w) \; dw = 0, \qquad a, \epsilon >0.
\end{equation}
Dividing both sides of (\ref{eq:cauchy:q}) by $2a$ and taking the limit $a\to \infty$, $\epsilon \to 0^+$,
we have 
\begin{equation}\label{eq:q:P0}
\hat{\vartheta}_{\bds 0} = P_0[\vartheta(\alpha)] = P_0[\vartheta(\alpha-ih)],
\end{equation}
where we use (\ref{eq:fhat:from:f}) in the first equality.
Substituting (\ref{eq:zt:zw:im:sfc}) and (\ref{eq:zt:zw:im:btm}) into (\ref{eq:q:P0}), we obtain the  
the time evolution equation of the width of the strip $S_{ h}$
\begin{equation}\label{eq:h:t}
 h_t = -P_0\bigg[\frac{\psi_\alpha^\sfc}{J^\sfc}\bigg].
\end{equation}
Finally, combining (\ref{eq:zt:by:zw:sfc}), (\ref{eq:zt:by:zw:btm}) and (\ref{eq:kinematic:hilb}),
we obtain the kinematic boundary conditions at both the free surface and the bottom boundary in
conformal space
\begin{equation}\label{eq:kinematic:conf}
\begin{pmatrix}
    \xi_t^\sfc \\[7pt]
    \eta_t^\sfc
  \end{pmatrix}
  = 
  \begin{pmatrix}
    \xi_\alpha^\sfc & -\eta_\alpha^\sfc\\[7pt]
    \eta_\alpha^\sfc & \xi_\alpha^\sfc
  \end{pmatrix}
  \begin{pmatrix}
    -H^{\coth}\left[\frac{\psi_\alpha^\sfc}{J^\sfc}\right] + C_1\\[7pt]
    -\frac{\psi_\alpha^\sfc}{J^\sfc}
  \end{pmatrix},
  \qquad
\begin{pmatrix}
    \xi_t^\btm \\[7pt]
    \eta_t^\btm
  \end{pmatrix}
   = \begin{pmatrix}
    \xi_\alpha^\btm \\[7pt]
    \eta_\alpha^\btm
  \end{pmatrix}
 \bigg(H^{\csch}\left[\frac{\psi_\alpha^\sfc}{J^\sfc}\right] + C_1\bigg).
\end{equation}
Since $\xi^\sfc$ and $\xi^\btm$ are determined by $\eta^\sfc$
and $\eta^\btm$ up to an additive constant $x_0$ by (\ref{eq:xi:eta:hilb}),
we only need to evolve $ h$, $\eta^\sfc$ and $\eta^\btm$
to track the evolution of the fluid domain.  
Comparing (\ref{eq:x:form}) and (\ref{eq:kinematic:conf}), we know that  the free parameter $x_0$
is related to $C_1$ through the ODE
\begin{equation}\label{eq:x0:evol}
  \frac{dx_0}{dt} = \Projection_0\left[
    \xi_\alpha^\sfc\left(-H^{\coth}\left[\frac{\psi_\alpha^\sfc}{J^\sfc}\right]
      + C_1\right) + \frac{\eta_\alpha^\sfc\psi_\alpha^\sfc}{J^\sfc}\right].
\end{equation}
Thus, $x_0(t)$ is uniquely determined by $C_1$ and $x_0(0)$.
Several choices of $C_1$ have been discussed in detail in \cite{quasi:ivp}. 
In the scope of this paper, we choose $C_1$ and $x_0(0)$ as follows
\begin{equation}
C_1=\Projection_0\big[\xi_\alpha^\sfc H^{\coth}[\psi_\alpha^\sfc/J^\sfc] -
      \eta_\alpha^\sfc\psi_\alpha^\sfc/J^\sfc\big], \qquad \qquad 
x_0(0) = 0.
\end{equation}
This ensures that $x_0(t) = 0$ for $t \ge 0$
and alleviates the need to explicitly solve the ODE \eqref{eq:x0:evol}.

Now we derive the dynamic boundary condition at the free surface in conformal space from (\ref{eq:Bernoulli}).
Differentiating the first equation in (\ref{eq:varphi:psi:bdr}) with respect to $t$, we obtain
\begin{equation}\label{eq:varphi:t:phys:conf}
\varphi_t^\sfc = \varphi_x^{\sfc, \ph} \xi_t^\sfc + \varphi_t^{\sfc, \ph},
\end{equation}
where $\varphi_x^{\sfc, \ph}$ can be expressed in terms of the gradient of $\Phi^\ph$ as follows
\begin{equation}\label{eq:varphi:x:Phi}
\varphi_x^{\sfc, \ph} = \Phi^\ph_x + \Phi^\ph_y\eta^{\sfc, \ph}_x.
\end{equation}
From (\ref{eq:potential:chain:rule}), we know that $\big|\nabla\Phi^\ph\big|^2=\left(\big(\varphi_\alpha^\sfc\big)^2+\big(\psi_\alpha^\sfc\big)^2\right)/J^\sfc$ at $\beta = 0$.
Substitution of (\ref{eq:kinematic:conf}), (\ref{eq:varphi:t:phys:conf}) and (\ref{eq:varphi:x:Phi}) into (\ref{eq:Bernoulli}) then gives 
\begin{equation}\label{eq:bernoulli:conf}
  \varphi_t^\sfc = \underbrace{\big(\Phi^\ph_x\,,\,\Phi^\ph_y\big)\begin{pmatrix}
    \xi_\alpha^\sfc & -\eta_\alpha^\sfc\\
    \eta_\alpha^\sfc & \xi_\alpha^\sfc
  \end{pmatrix}}_{\jd\big(\varphi_\alpha^\sfc\,,\,-\psi_\alpha^\sfc\big)}
  \begin{pmatrix}
    -\Hilbert^{\coth}\left[\psi_\alpha^\sfc/J^\sfc\right] + C_1\\
    -\psi_\alpha^\sfc/J^\sfc
  \end{pmatrix}
  - \frac{\big(\varphi_\alpha^\sfc\big)^2+\big(\psi_\alpha^\sfc\big)^2}{2J^\sfc} - g\eta^\sfc + \tau\kappa + C,
\end{equation}
where $\kappa$ is the mean curvature, given by
\begin{equation}\label{eq:kappa:conf}
\kappa = \frac{\xi^\sfc_\alpha\eta^\sfc_{\alpha\alpha} - 
\eta^\sfc_\alpha\xi^\sfc_{\alpha\alpha}}{\big(J^\sfc\big)^{3/2}},
\end{equation}
and $C$ is an arbitrary integration constant that may depend on time
but not space. In the discussion of this paper, we choose $C$ such
that $P_0[\varphi^\sfc_t] = 0$.  In conclusion, (\ref{eq:h:t}),
(\ref{eq:kinematic:conf}) and (\ref{eq:bernoulli:conf}) are the
governing equations in conformal space for finite-depth quasi-periodic
gravity-capillary waves.

Following \cite{quasi:ivp}, instead of solving these equations
directly, which are posed on the real line, we lift the problem to a
higher dimensional torus $\mbb T^d$ and compute the time evolution of
the corresponding torus functions; then we evaluate the torus
functions along the characteristic direction to obtain quasi-periodic
functions on the real line.  Using the torus version of the
quasi-periodic derivative and Hilbert transform operators in
Definitions \ref{def:deriv} and \ref{def:hilbert}, we obtain the
governing equations on $\mbb T^d$ from (\ref{eq:h:t}),
(\ref{eq:kinematic:conf}) and (\ref{eq:bernoulli:conf}),
\begin{equation} \label{eq:time:evol:torus}
\begin{gathered}
\td\eta_t^\sfc = \big({-}\Hilbert^{\coth}[\td \chi^\sfc] +C_1\big)\td\eta_\alpha^\sfc
- \td\xi_\alpha^\sfc\td\chi^\sfc,  
\qquad 
\td\eta_t^\btm = \big(H^{\csch}[\td\chi^\sfc]+ C_1\big)\td\eta_\alpha^\btm,
\qquad
 h_t = -P_0[\td \chi^\sfc],
\\
 \td\varphi_t^\sfc = P\bigg[\frac{\big(\td\psi_\alpha^\sfc\big)^2 - \big(\td\varphi_\alpha^\sfc + \mc U\big)^2}{2\td J^\sfc} +
 \big(C_1-\Hilbert^{\coth}[\td\chi^\sfc] \big)\big(\td\varphi_\alpha^\sfc+\mc U\big) - g\td\eta^\sfc +
    \tau\td\kappa\bigg]
\\
\td\xi^\sfc =  H^{\coth}[\td\eta^\sfc] 
+ H^{\csch}[\td\eta^\btm], \qquad \qquad  
  \td\psi^\sfc = H^{\tanh}[\td\varphi^\sfc], 
\\
\td J^\sfc= \big(1+\td \xi_\alpha^\sfc\big)^2 + \big(\td\eta_\alpha^\sfc\big)^2, 
\qquad \qquad 
\td\chi^\sfc= \frac{\td\psi_\alpha^\sfc}{\td J^\sfc},
 \\
\td\kappa = \frac{\big(1+\td\xi_\alpha^\sfc\big)\td\eta_{\alpha\alpha}^\sfc -
      \td\eta_\alpha^\sfc\td\xi_{\alpha\alpha}^\sfc}{(\td J^\sfc)^{3/2}},
\qquad \qquad
C_1=P_0\big[\big(1+\td\xi_\alpha^\sfc\big) H^{\coth}[\td\chi^\sfc] -
      \td\eta_\alpha^\sfc\td\chi^\sfc\big].
\end{gathered}
\end{equation}
We remark that $\td \varphi$, which is defined on $\mbb T^d$,
represents only the quasi-periodic part of $\varphi$. An extra term $\mc U\alpha$ is included in the definition (\ref{eq:varphi:sfc})
to account for the background flow (when present). Similarly, $\xi^\sfc$ and $\xi^\btm$ are obtained from $\td \xi^\sfc$ and $\td \xi^\btm$ via 
\begin{equation}\label{eq:xi:tilde}
\xi^\sfc(\alpha, t) = \alpha + \td\xi^\sfc(\bds k\alpha, t), \qquad \qquad
\xi^\btm(\alpha, t) = \alpha + \td\xi^\btm(\bds k\alpha, t),
\end{equation}
where $\td \xi^\sfc$ is given in (\ref{eq:time:evol:torus}) and $\td \xi^\btm$ is given by
\begin{equation}
\td\xi^\btm =  - H^{\csch}[\td\eta^\sfc] - H^{\coth}[\td\eta^\btm].
\end{equation}
According to (\ref{eq:xi:tilde}), we have
\begin{equation}
\xi^\sfc_\alpha(\alpha, t) = 1 + \td\xi^\sfc_\alpha(\bds k\alpha, t), \qquad \qquad
\xi^\btm_\alpha(\alpha, t) = 1 + \td\xi^\btm_\alpha(\bds k\alpha, t),
\end{equation}
which is the reason $(1+\td\xi^\sfc_\alpha)$ appears in various places
in (\ref{eq:time:evol:torus}). We also note that the operators
  $H^{\coth}$, $H^{\csch}$ and $H^{\tanh}$ in
  \eqref{eq:time:evol:torus} vary in time along with $h(t)$.

\begin{remark}
 Modifying the analysis of
 \cite{quasi:ivp}, one can show that if $\zeta^\sfc$ and $\zeta^\btm$ are injective, then $\eta^{\sfc, \ph}$ and $\eta^{\btm, \ph}$ are 
also quasi-periodic functions of the same quasi-periods. Moreover, the corresponding torus functions 
$\td \eta^{\sfc, \ph}$ and $\td \eta^{\btm, \ph}$ can be obtained from $\td \eta^\sfc$ and $\td \eta^\btm$ by
\begin{equation}\label{eq:A}
\td \eta^{\sfc, \ph}(\bds x, t) = \td \eta^\sfc(\bds x + \bds k \td {\mc A}^\sfc(\bds x, t), t),  \qquad 
\td \eta^{\btm, \ph}(\bds x, t) = \td \eta^\btm(\bds x + \bds k \td {\mc A}^\btm(\bds x, t), t),
\end{equation}
where $\td{\mc A}^\sfc$ and $\td{\mc A}^\btm$ satisfy 
\begin{equation}
\td{\mc A}^\sfc(\bds x, t) + \td \xi^\sfc(\bds x + \bds k \td {\mc A}^\sfc(\bds x, t), t) = 0, \qquad
\td{\mc A}^\btm(\bds x, t) + \td \xi^\btm(\bds x + \bds k \td {\mc A}^\btm(\bds x, t), t) = 0.
\end{equation}
In numerical computations, at any given $t$, one can formulate (\ref{eq:A}) as a nonlinear least-squares problem and solve it using a Levenberg-Marquardt method \cite{wilkening2012overdetermined}, which is discussed in Appendix \ref{sec:conformal:transform}.
\end{remark}
\begin{remark}\label{rmk:mu:def}
  Since the bottom boundary is stationary, conservation of mass requires that
  the mean surface height, which we denote by
\begin{equation} \label{eq:mu:def}
  \mu =
  \frac1{(2\pi)^d}\int_{\mbb T^d} \td\eta^{\sfc,\ph}\,dx_1\cdots dx_d =
  \frac1{(2\pi)^d}\int_{\mbb T^d} \td\eta^\sfc(1+\td\xi^\sfc_\alpha)\,d\alpha_1\cdots d\alpha_d,
\end{equation}
is a constant in time. Indeed, one finds that $\mu_t=0$ by
differentiating the second formula of \eqref{eq:mu:def} under the
integral sign, integrating the term $\td\eta^\sfc\td\xi^\sfc_{\alpha t}$ by
parts with respect to $\alpha$, and using
\eqref{eq:kinematic:sfc}, keeping in mind that
  $\xi^\sfc_\alpha=1+\td\xi^\sfc_\alpha$ due to \eqref{eq:xi:tilde}.
  One usually assumes $\mu=0$, though for
traveling waves it is convenient to first compute the wave assuming
$\hat\eta_{\bds 0}^\sfc=0$ and then adjust $\hat\eta^\sfc_{\bds 0}$ and
$\hat\eta^\btm_{\bds 0}$ at the end to achieve $\mu=0$.
\end{remark}

\begin{remark}\label{rmk:flat:btm}
  The governing equations (\ref{eq:time:evol:torus}) still hold when
  the bottom boundary of the fluid domain is flat: $\eta^{\btm,
    \ph}(x) = -h^\ph$. In the usual case that $\mu=0$, $h^\ph$ is the
  mean depth of the fluid in physical space. Otherwise the mean
  depth is $\mu+h^\ph$. From \eqref{eq:h:free:param}, we have
  \begin{equation}\label{eq:flat:hat:eta:0}
    h^\ph = -\eta^\btm = -\hat\eta^\btm_{\bds 0} =
    h - \hat\eta^\sfc_{\bds 0},
  \end{equation}
  which is a constant independent of $\alpha$ and $t$
  even though $h$ and $\hat\eta^\sfc_{\bds 0}$ vary in
  time. Moreover, $\xi^\sfc$ is related to $\eta^\sfc$  by 
\begin{equation}\label{eq:xi:eta:hilb:flat}
\xi^\sfc = \alpha + H^{\coth}[\eta^\sfc].
\end{equation}
Therefore, when the bottom boundary is flat, one only needs to evolve
$\td\eta^\sfc$, $\td\varphi^\sfc$ and $ h$.
\end{remark}

\begin{remark}\label{rmk:periodic}
  Even though we derive (\ref{eq:time:evol:torus}) in the quasi-periodic setting, these equations still hold for the periodic problem if we set $d=1$ and $\bds k = (1)$.
To obtain the governing equations on $\mbb T$, one just needs to replace the quasi-periodic Hilbert transforms by their periodic counterparts in (\ref{eq:time:evol:torus}), which can be obtained by changing $\la\bds j, \bds k \ra$ to $j$ in (\ref{eq:H:symbol}). If $d>1$, the periodic problem may
be embedded in the quasi-periodic problem by assuming that each of the
torus functions in \eqref{eq:time:evol:torus} is independent of
$\alpha_2,\dots,\alpha_d$.
\end{remark}

\section{Quasi-periodic traveling waves } \label{sec:traveling}


\subsection{Governing equations of quasi-periodic traveling waves}
\label{sec:govern:eq:quasi:trav}
For traveling waves, the system should be translation invariant, so we
assume the bottom boundary is flat. According to
Remark~\ref{rmk:flat:btm}, we only need to consider the surface
variables in this case. Thus, to simplify the notation, we drop the
superscript ``$\sfc$'' in these variables in this section. Moreover,
as discussed in Section \ref{sec:govern:phys}, we focus our discussion
on the laboratory frame and assume that there is no background flow.

Since the bottom boundary is flat, $\xi$, $\eta$ and
$\varphi$, $\psi$ are related by Hilbert transforms
\begin{equation}\label{eq:trav:hilb}
\xi = \alpha + H^{\coth}[\eta], \qquad 
\xi_\alpha = 1 + H^{\coth}[\eta_\alpha], \qquad 
\varphi = H^{\coth}[\psi], \qquad
\varphi_\alpha = H^{\coth}[\psi_\alpha].
\end{equation}
We assume the wave is traveling from left to right at speed $c$; therefore, we have 
\begin{equation}\label{eq:trav:init:phys}
\eta^\ph(x, t) = \eta_0^\ph(x-ct), \qquad \qquad
\varphi^\ph(x, t) = \varphi_0^\ph(x-ct).
\end{equation}
Differentiating both sides of  (\ref{eq:trav:init:phys}) with respect to $x$ and $t$ separately, 
we know that a traveling solution satisfies
\begin{equation}\label{eq:eta:varphi:driv:trav:phy}
\eta^\ph_t = -c\eta^\ph_x, \qquad \qquad
\varphi^\ph_t = -c\varphi^\ph_x.
\end{equation}
Substituting the second equation of (\ref{eq:deta:sfc}) into the first equation of (\ref{eq:eta:varphi:driv:trav:phy})
and multiplying both sides of the equation by $\xi^\sfc_\alpha$, we obtain
\begin{equation}\label{eq:kinematic:trav}
\eta_t\xi_\alpha - \xi_t \eta_\alpha = -c\eta_\alpha.
\end{equation}
Comparing (\ref{eq:kinematic:trav}) and (\ref{eq:kinematic:sfc}), we conclude that a traveling solution satisfies
\begin{equation}\label{eq:eta:psi:trav:conf}
\psi_\alpha = c\eta_\alpha 
\end{equation}
in conformal space. Applying the Hilbert transform $H^{\coth}$ to both sides of (\ref{eq:eta:psi:trav:conf}), we obtain 
\begin{equation}\label{eq:xi:varphi:trav:conf}
\varphi_\alpha = c(\xi_\alpha-1).
\end{equation}
Substituting the traveling condition of $\varphi^\ph$ in (\ref{eq:eta:varphi:driv:trav:phy}) into
(\ref{eq:varphi:t:phys:conf}) and employing (\ref{eq:varphi:x:Phi}) to express $\varphi^\ph_x$ in terms of the gradient of $\Phi^\ph$,
we obtain that 
\begin{equation}\label{eq:varphi:t:trav:conf}
\begin{aligned}
\varphi_t &= \big(\Phi^\ph_x + \Phi^\ph_y \eta^\ph_x\big)(\xi_t - c)
= \frac{\varphi_\alpha}{\xi_\alpha}
\big(\xi_\alpha\big(-H^{\coth}\Big[\frac{\psi_\alpha}{J}\Big] + C_1\big) + \eta_\alpha \frac{\psi_\alpha}{J} - c\big) \\
& =  \frac{\varphi_\alpha}{\xi_\alpha}
\bigg(\xi_\alpha\big(-H^{\coth}\Big[\frac{\psi_\alpha}{J}\Big] + C_1\big) 
+ \frac{c\big(\eta_\alpha\big)^2}{J} - c\bigg)
 = \varphi_\alpha \bigg(-H^{\coth}\Big[\frac{\psi_\alpha}{J}\Big]+ C_1 - \frac{c\xi_\alpha}{J}\bigg).
\end{aligned}
\end{equation}
Here in the second equality, we use the first equation in (\ref{eq:deta:sfc}) to rewrite $\eta^\ph_x$ as
$\eta^\sfc_\alpha/\xi^\sfc_\alpha$ and substitute the gradient of $\Phi^\ph$ and $\xi_t$ using 
(\ref{eq:velocity:sfc}) and (\ref{eq:kinematic:conf}), respectively.
In the third equality, we use (\ref{eq:eta:psi:trav:conf}) to replace $\psi_\alpha$ by $c\eta_\alpha$.
The substitution of (\ref{eq:eta:psi:trav:conf}) and (\ref{eq:varphi:t:trav:conf}) into (\ref{eq:bernoulli:conf})
gives 
\begin{equation}\label{eq:trav:conf:1}
\frac{c}{J}\big(\varphi_\alpha \xi_\alpha + \psi_\alpha \eta_\alpha\big)
 - \frac{1}{2J}\big((\varphi_\alpha)^2 + (\psi_\alpha)^2\big)
 -g \eta + \tau \kappa + C = 0.
\end{equation}
Using (\ref{eq:eta:psi:trav:conf}) and (\ref{eq:xi:varphi:trav:conf}) to express $\varphi_\alpha$ and $\psi_\alpha$ in terms of $\xi_\alpha$ and $\eta_\alpha$, respectively, we obtain the governing equation of traveling waves
\begin{equation}\label{eq:trav:conf}
P\bigg[\frac{c^2}{2J} + g\eta -\tau\kappa\bigg] = 0,
\end{equation}
where we choose the integration constant $C$ in \eqref{eq:trav:conf:1} such that 
$P_0$ acting on the left-hand side of (\ref{eq:trav:conf:1}) returns zero.
Since (\ref{eq:trav:conf}) does not depend on time, the solution of (\ref{eq:trav:conf}) can be considered as the initial condition of a traveling wave. 
From (\ref{eq:trav:hilb}) and (\ref{eq:kappa:conf}),
we know that $J$ and $\kappa$ are determined by $\eta$;  
hence, the unknowns in (\ref{eq:trav:conf}) are $\tau$, $c$ and $\eta$.
Even though we are mainly interested in the case where $\eta$ is quasi-periodic,
the governing equation (\ref{eq:trav:conf}) still holds when $\eta$ is
periodic.  Due to the projection operator, modifying $\eta$ by a
constant will not influence (\ref{eq:trav:conf}); hence, we assume
that $P_0[\eta] = 0$.  In this paper, we focus on traveling waves with
even symmetry
\begin{equation}
\eta(\alpha) = \eta(-\alpha).
\end{equation}
We compute $\xi$ from $\eta$ using (\ref{eq:trav:hilb}) and deduce that $\xi$ is odd.
Asymmetric traveling waves have been studied in \cite{wang2014asymmetric, gao2016asymmetric, zufiria:1987:symmetry:breaking} in the periodic setting. 

As in the initial value problem, 
we first solve for $\td \eta$ on $\mbb T^d$ and then reconstruct $\eta$ from $\td \eta$ using (\ref{eq:eta:sfc:btm:quasi}).
The governing equations of traveling waves on the torus read
\begin{equation}\label{eq:trav:torus:conf}
\begin{gathered}
\mc R[\tau, b, \td \eta]  = P\bigg[ \frac{b}{2\td J} + g\td\eta  - \tau\td\kappa\bigg] = 0,\\ 
\td\xi = H^{\coth}[\td\eta], \qquad\qquad
    \td J = \big(1+\td\xi_\alpha\big)^2 + \td\eta_\alpha^2, \qquad\qquad
   \td \kappa = \frac{\big(1+\td\xi_\alpha\big)\td\eta_{\alpha\alpha} -
      \td\eta_\alpha\td\xi_{\alpha\alpha}}{\td J^{3/2}}, 
\end{gathered}
\end{equation} 
where $b = c^2$ and $\mc R$ is called the residual function. We treat the strip width
$h$ in conformal space as a fixed parameter and suppress it in the argument list of
$\mc R$; see Remark~\ref{rmk:mean:quasi} below.
Linearizing (\ref{eq:trav:torus:conf}) around the zero solution $\td\eta = 0$, we obtain
\begin{equation}\label{eq:linear:trav}
bH^{\coth}[\delta \tilde \eta_\alpha] - g\delta{\tilde \eta} +
  \tau  \delta \tilde\eta_{\alpha\alpha} = 0,
\end{equation}
where $\delta \tilde\eta $ denotes the variation of $\tilde \eta$. 
Expressing $\delta \tilde\eta$ in terms of its Fourier series in (\ref{eq:linear:trav}), we obtain the dispersion relation for the linearized problem
\begin{equation}\label{eq:dispersion:quasi}
b\coth(\la \bds j, \bds k \ra h) \la \bds j, \bds k \ra- g - \tau (\la \bds j, \bds k\ra)^2 = 0, 
\qquad \qquad \bds j \in \mbb \mbb Z^d.
\end{equation}
Since the entries of $\bds k$ are linearly independent over $\mbb{Z}$, given $b$ and $\tau$, 
there exist at most two linearly independent vectors $ \bds j_1$, $ \bds j_2 \in \mbb{Z}^d$ 
that satisfy the dispersion relation \cite{bridges1996spatially}. 
For simplicity, we consider the basic case where $d = 2$; hence, $\eta$ possesses two quasi-periods and $\td \eta$ is defined on $\mbb T^2$. 
Without loss of generality, we also assume that 
$\bds{j}_1 = (1,0)^T$, $\bds{j}_2 = (0,1)^T$ and
$\bds k  = (1, k)^T$, where $k$ is a positive irrational number.

In summary, we study quasi-periodic traveling waves of the following form
\begin{equation}\label{eq:eta:two:quasi}
\eta (\alpha) = \tilde{\eta}(\alpha, k\alpha), 
\qquad \qquad 
\tilde{\eta}(\alpha_1, \alpha_2) = \sum_{j_1, j_2 \in \mbb Z} 
\hat{\eta}_{j_1, j_2} e^{i(j_1 \alpha_1 + j_2\alpha_2)}.
\end{equation}
We also assume that $\td \eta $ is an even function with zero mean on $\mbb T^2$ in conformal space,
which is consistent with the assumptions on $\eta$.
Therefore the Fourier coefficients of $\tilde \eta$ satisfy
\begin{equation}\label{eq:eta:assump}
\hat{\eta}_{0, 0} = 0, \qquad \qquad \hat{\eta}_{j_1, j_2} = \hat{\eta}_{-j_1, -j_2}\in\mbb R.
\end{equation}
We refer to Remark \ref{rmk:mean:quasi}
below if one wants to obtain solutions with zero mean in physical space.
Under assumptions (\ref{eq:eta:two:quasi}) and (\ref{eq:eta:assump}), we can study the problem of quasi-periodic traveling waves in the setting of a bifurcation problem with a two-dimensional kernel
spanned by the solutions of the linearized problem (\ref{eq:linear:trav}):
\begin{equation} \label{eq:linear_solution}
\begin{gathered}
\tilde\eta_{\text{lin}}(\alpha_1,\alpha_2) =
\hat\eta_{1,0}(e^{i\alpha_1}+e^{-i\alpha_1})
+ \hat\eta_{0,1}(e^{i\alpha_2}+e^{-i\alpha_2}), \\ 
b_{\text{lin}} = c_{\text{lin}}^2 =  \frac{g(k^2-1)}{k(k\coth(h) - \coth(kh))}, \qquad \qquad
\tau_{\text{lin}}=  \frac{g(k\coth(kh)-\coth(h))}{k(k\coth(h)-\coth(kh))}.   
\end{gathered}
\end{equation}
We refer to $\hat{\eta}_{1,0}$ and $\hat{\eta}_{0,1}$ as the base Fourier coefficients 
and the corresponding Fourier modes $e^{\pm i\alpha_1}$, $e^{\pm i\alpha_2}$
as the base Fourier modes.
Nonlinear solutions can be considered as 
bifurcations from the zero-amplitude solution.
We usually choose the base Fourier coefficients 
as bifurcation parameters and fix them at nonzero values to ensure that the solutions we obtain
are genuinely quasi-periodic. In finite depth, $h$ is a third parameter.

As shown in \cite{quasi:bifur}, large-amplitude quasi-periodic
traveling solutions can often be found by searching for secondary
bifurcations from finite-amplitude periodic traveling waves. The
linearization of (\ref{eq:trav:torus:conf}) around a periodic solution
reads
\begin{equation}\label{eq:variational:eq}
  \begin{gathered}
   \delta{\mathcal{R}} = P\left[\frac{\delta b}{
        2\tilde J}  -
      \frac{1}{2\tilde J^2} b \delta{\td J} + g\delta{\td\eta} -
      \delta\tau\tilde\kappa - \tau\delta{\td\kappa}\right],\\ 
    \delta{\td\xi}_\alpha = H ^{\coth}[\delta{\td\eta}_\alpha], \qquad\quad
    \delta{\td J} = 2\left((1+\td\xi_\alpha)\delta{\td\xi}_\alpha  +
      \td\eta_\alpha\delta{\td\eta}_\alpha \right), \\ 
    \delta{\td\kappa} =  -\frac{3}{2}\td\kappa\frac{\delta \td J}{\td J}
    + \frac{1}{\td J^{3/2}}
    \Big(\delta{\td\xi}_\alpha\td\eta_{\alpha\alpha}+
      (1+\td\xi_\alpha)\delta{\td \eta}_{\alpha\alpha}-
      \delta{\td \eta}_\alpha \td\xi_{\alpha\alpha}-
      \td\eta_{\alpha}\delta{\td \xi}_{\alpha\alpha}\Big).
  \end{gathered}
\end{equation}
Let $q$ denote the triple $(\tau,b,\td\eta)$ and let $\mfq^\per(s)$
denote a one-parameter family of periodic traveling waves embedded in
the quasi-periodic framework by assuming $\td\eta(\alpha_1,\alpha_2)$
is independent of $\alpha_2$. Here $s$ is an amplitude parameter (such
  as $\hat\eta_{1,0}$), and, for simplicity, we fix $\tau$ and the
strip width $h$ in conformal space to be independent of $s$. Each
solution $q=\mfq^\per(s)$ in the family satisfies $\mc
R\big(q\big)=0$. In \cite{quasi:bifur}, an algorithm is presented for
locating bifurcation points by using a quadratically convergent
root bracketing technique \cite{brent:73} to locate zeros of
the signed smallest singular value
\begin{equation}\label{eq:chi:def}
  \chi(s) = \sgn\Big( \det\Big( \mc J^\qua(s) \Big) \Big) \sigma_\text{min}\Big( \mc J^\qua(s) \Big).
\end{equation}
Here $\mc J^\qua(s)$ is a Fourier truncation of the restricted
Jacobian obtained from the linearization \eqref{eq:variational:eq}
applied only in quasi-periodic perturbation directions of the form
$\delta q=(0,0,\delta\td\eta^\qua)$, where $\delta\td\eta^\qua$ has 2D
Fourier modes $\widehat{\delta\td\eta}^\qua_{j_1,j_2}$ that are
all zero unless $j_2\in\{1,-1\}$. This construction is based on
Bloch-Fourier perturbation theory over periodic potentials
\cite{kittel:book}. At zeros of $\chi(s)$, $\mc J^\qua(s)$ has a
kernel that provides a bifurcation direction $\delta\td\eta^\qua$ that
allows us to switch from the primary periodic branch to the secondary
quasi-periodic branch of traveling waves. We use $\td\eta^\per +
\epsilon \delta \td\eta^\qua$, with $\epsilon$ chosen empirically, as
an initial guess for solutions on this secondary branch, and then use
numerical continuation to follow the branch beyond the realm of
linearization about the primary branch. Further discussion of the
analysis and computation of the bifurcation problem in the
infinite-depth setting is given in \cite{quasi:bifur}.

\begin{remark}\label{rmk:mean:quasi}
We have simplified the computation of quasi-periodic traveling waves
via the conformal mapping formulation by setting $\hat{\eta}_{0,0}=0$
and fixing the strip width $h$ in conformal space. By
  \eqref{eq:flat:hat:eta:0}, this causes the vertical position of the
  bottom boundary in physical space to be $\eta^\btm=-h^\ph=-h$. However, the
  mean surface height $\mu$ in \eqref{eq:mu:def} is generally non-zero
  when $\hat\eta_{0,0}=0$, so the physical fluid depth is $\mu+h$.
If desired, after computing a solution with $\hat{\eta}_{0,0}=0$, one
can compute $\mu$ via \eqref{eq:mu:def} and shift the vertical
  position of both the free surface and bottom boundary by $-\mu$ in
  physical space. This will not change the parameter $h$, so it will
  still be a traveling wave. When $\mu$ and $h^\ph$ are computed for
  the new wave, the former will be zero and the latter will be the
  physical fluid depth. This shifted solution satisfies
\begin{equation}\label{eq:eta:0:mean}
  \hat{\eta}_{0,0} = -P_0[ \big(P[\tilde\eta]\big)(1+\tilde \xi_\alpha)].
\end{equation}
Another option is to prescribe
$\mu=0$, $h^\ph$, $\hat\eta_{1,0}$ and $\hat\eta_{0,1}$ and solve for
$\hat{\eta}_{0,0}$ and $h$ along with the remaining Fourier modes
$\hat\eta_{j_1,j_2}$ using the Levenberg-Marquardt solver. This
would entail including $h=h^\ph+\hat\eta_{0,0}$ from
\eqref{eq:flat:hat:eta:0} as well as \eqref{eq:eta:0:mean} as
additional constraints in \eqref{eq:trav:torus:conf}.
\end{remark}


\subsection{Weakly nonlinear approximations of quasi-periodic traveling waves}
\label{sec:weak:approx}
Although the primary focus of this work is on computing quasi-periodic
solutions of the fully nonlinear time-dependent and traveling water
wave equations in finite depth, it is instructive to investigate how
small divisors arise in weakly nonlinear approximations of
small-amplitude quasi-periodic traveling waves. In previous work, it
has been necessary to treat such small divisors carefully using
Nash-Moser theory \cite{plotnikov01,iooss05} to prove existence of
temporally quasi-periodic water waves
\cite{berti2016quasi,baldi2018time,berti2021traveling,berti2021pure,feola2020trav}.
Here we focus on spatial quasi-periodicity.

As discussed in Section \ref{sec:govern:eq:quasi:trav}, the traveling
solutions bifurcating from the zero solution form a three-parameter
family with bifurcation parameters $\hat{\eta}_{1,0}$,
$\hat{\eta}_{0,1}$ and $h$.  In the weakly nonlinear model, we treat $h$
as a constant and set these two
Fourier coefficients to be fixed, non-zero multiples of an amplitude
parameter $\veps$ and aim to express $b$, $\tau$ and the other Fourier
coefficients of $\td \eta$ in terms of them.  Let us consider the
following asymptotic expansions of $b$, $\tau$ and $\td \eta$
\begin{equation}\label{eq:asym:expansion}
\begin{aligned}
b = & b^{(0)} + \epsilon b^{(1)} + \epsilon^2 b^{(2)} + \epsilon^3 b^{(3)} + O(\epsilon^4), \\
\tau = & \tau^{(0)} + \epsilon \tau^{(1)} + \epsilon^2 \tau^{(2)} + \epsilon^3 \tau^{(3)} + O(\epsilon^4), \\
\tilde{\eta} = &\epsilon \tilde{\eta}^{(1)} + \epsilon^2 \tilde{\eta}^{(2)} + \epsilon^3 \tilde{\eta}^{(3)} + O(\epsilon^4).
\end{aligned}
\end{equation}
Substituting (\ref{eq:asym:expansion}) into 
(\ref{eq:trav:torus:conf}) and eliminating the coefficients of 
$\epsilon^n$ for $n = 0, 1, 2$, we obtain
\begin{align} \label{eq:asym:epsilon}
 {\scriptstyle O(1):} &\,{\scriptstyle P\bigg[\frac{1}{2} b ^{(0)}\bigg] = 0,} \notag \\ 
{\scriptstyle O(\epsilon):} &\, {\scriptstyle P\bigg[\frac{1}{2}b^{(1)} + g\tilde{\eta}^{(1)} 
- b^{(0)} H^{\coth}\big[\tilde\eta^{(1)}_\alpha\big]- 
\tau^{(0)}\tilde{\eta}^{(1)}_{\alpha\alpha}\bigg] = 0,} \\ 
{\scriptstyle O(\epsilon^2):} &\,
{\scriptstyle P\bigg[\frac{1}{2}b^{(2)} + g\tilde{\eta}^{(2)} - b^{(0)}H^{\coth}\big[\tilde{\eta}^{(2)}_\alpha\big] 
- \tau^{(0)} \td \eta^{(2)}_{\alpha\alpha}
- b^{(1)} H^{\coth}\big[\tilde{\eta}^{(1)}_\alpha\big] - \tau^{(1)}\tilde{\eta}^{(1)}_{\alpha\alpha}} \notag \\
&{\scriptstyle \qquad + b^{(0)}\left(\frac{3}{2} \left(H^{\coth}\big[\tilde{\eta}^{(1)}_\alpha\big]\right)^2 
- \frac{1}{2} \big(\tilde{\eta}^{(1)}_\alpha\big)^2\right) 
+\tau^{(0)}\left(2H^{\coth}\big[\tilde{\eta}^{(1)}_\alpha\big]\tilde{\eta}^{(1)}_{\alpha\alpha}
+ H^{\coth}\big[\tilde{\eta}^{(1)}_{\alpha\alpha}\big]\tilde{\eta}^{(1)}_\alpha\right) \bigg] = 0.} \notag
\end{align}
Since the constant term in (\ref{eq:asym:epsilon}) vanishes under the projection, 
the second equation is essentially the same as the linearization (\ref{eq:linear:trav}); therefore, we have
\begin{equation}\label{eq:eta:tau:b:order:1}
\tilde{\eta}^{(1)} = \td \eta_{\text{lin}} = \hat{\eta}_{1,0}e^{i\alpha_1} + \hat{\eta}_{0,1}e^{i\alpha_2} + c.c., \qquad 
b^{(0)} = b_{\text{lin}}, \qquad 
\tau^{(0)} = \tau_{\text{lin}}.
\end{equation}
Using the property of the projection operator and the assumption that $P_0[\tilde \eta] =0$, we rewrite 
the third equation in (\ref{eq:asym:epsilon}) as
\begin{align}  \label{eq:asym:order:2}
&\underbrace{g\tilde{\eta}^{(2)} - b^{(0)}H^{\coth}\big[\tilde{\eta}^{(2)}_\alpha\big] 
- \tau^{(0)}\tilde{\eta}^{(2)}_{\alpha\alpha}}_{A^{(2)}}
\underbrace{- b^{(1)} H^{\coth}\big[\tilde{\eta}^{(1)}_\alpha\big] 
- \tau^{(1)}\tilde{\eta}^{(1)}_{\alpha\alpha}}_{B^{(2)}}\\
= &\underbrace{P\left[
 b^{(0)}
 \left(-\frac{3}{2} \left(H^{\coth}\big[\tilde{\eta}^{(1)}_\alpha\big]\right)^2 
  + \frac{1}{2} \big(\tilde{\eta}^{(1)}_\alpha\big)^2\right)
-\tau^{(0)}
\left(2H^{\coth}\big[\tilde{\eta}^{(1)}_\alpha\big]\tilde{\eta}^{(1)}_{\alpha\alpha}
+ H^{\coth}\big[\tilde{\eta}^{(1)}_{\alpha\alpha}\big]\tilde{\eta}^{(1)}_{\alpha}\right)\right]}_{C^{(2)}}.
\notag 
\end{align}
Substituting $\td \eta^{(1)}$, $\td b^{(0)}$ and $\td \tau^{(0)}$ into $C^{(2)}$ 
using (\ref{eq:eta:tau:b:order:1}), we obtain
\begin{equation}
C^{(2)} = \hat{C}_{2,0}^{(2)}e^{i(2\alpha_1)} + \hat{C}_{0,2}^{(2)}e^{i(2\alpha_2)}
+ \hat{C}_{1,1}^{(2)}e^{i(\alpha_1 + \alpha_2)} + \hat{C}_{1,-1}^{(2)}e^{i(\alpha_1-\alpha_2)}
+ c.c., 
\end{equation}
where the Fourier coefficients of $C^{(2)}$ are 
\begin{equation}\label{eq:C:2}
{\scriptsize
\begin{aligned}
\hat{C}_{2,0}^{(2)} &= g\hat{\eta}_{1,0}^2 
\frac{3(k^2+1)\coth^2(h) - 6k\coth(kh)\coth(h) + k^2-1}
{2k(\coth(kh)-k\coth(h))}, \\ 
\hat{C}_{0,2}^{(2)} &= -gk\hat{\eta}_{0,1}^2
\frac{3(k^2+1)\coth^2(kh) - 6k\coth(kh)\coth(h)-k^2+1}
{2(\coth(kh)-k\coth(h))},\\ 
\hat{C}_{1,1}^{(2)} &= -g\hat{\eta}_{1,0}\hat{\eta}_{0,1}
\frac{(k^2+2k)\coth^2(kh) - (2k+1)\coth^2(h)+ (-k^2+1)\coth(kh)\coth(h) - k^2+1}
{\coth(kh)- k\coth(h)}, \\ 
\hat{C}_{1,-1}^{(2)} &= g\hat{\eta}_{1,0}\hat{\eta}_{0,1}
\frac{(k^2-2k)\coth^2(kh) + (2k-1)\coth^2(h) + (k^2-1)\coth(kh)\coth(h) - k^2+1}
{\coth(kh)-k\coth(h)}.
\end{aligned}}
\end{equation}

We observe that $A^{(2)}$ is linear with respect to $\tilde{\eta} ^{(2)}$ and 
the Fourier coefficients of $A^{(2)}$ can be expressed as
\begin{equation}\label{eq:A:order:2}
\hat{A}_{j_1, j_2}^{(2)} = \hat{S}_{j_1, j_2}\hat{\eta}^{(2)}_{j_1, j_2},
\end{equation}
where the symbol $\hat{S}_{j_1, j_2}$ is defined by
\begin{align}\label{eq:S}
{\scriptstyle
\hat{S}_{j_1, j_2}}
&{\scriptstyle = g - b^{(0)}\coth((j_1 + kj_2)h)(j_1 + kj_2) + \tau^{(0)}(j_1 + k j_2)^2}\\
&{\scriptstyle = \frac{g}{k}\left(k + \frac{k^2-1}{ \coth(kh)- k\coth(h)}\coth((j_1 + kj_2)h)(j_1 + k j_2)
+ \frac{\coth(h)-k\coth(kh)}{\coth(kh)-k\coth(h)}(j_1 + k j_2)^2\right).} \notag
\end{align}
Since $\hat{S}_{\pm1,0}$ and $\hat{S}_{0,\pm1}$ are both zero according to the definition,
we know that $\hat{A}^{(2)}_{\pm1, 0} = \hat{A}^{(2)}_{0, \pm1} = 0$.
We also observe that $B^{(2)}$ is linear with respect to $\td \eta^{(1)}$ with Fourier coefficients
\begin{equation}\label{eq:B:2}
\hat B^{(2)}_{j_1, j_2} =  \hat{Q}_{j_1, j_2}^{(1)}\hat{\eta}^{(1)}_{j_1, j_2}, \qquad 
\hat{Q}_{j_1, j_2}^{(n)} = -b^{(n)}\coth((j_1 + kj_2)h)(j_1 + k j_2) + \tau^{(n)}(j_1 +kj_2)^2,
\end{equation}
where $(j_1, j_2) = (\pm1,0), (0,\pm 1)$ according to (\ref{eq:eta:tau:b:order:1}).
Combining (\ref{eq:C:2}), (\ref{eq:A:order:2}) and (\ref{eq:B:2}), we obtain 
\begin{equation}
b^{(1)} = \tau^{(1)} = 0, \qquad \qquad
\hat{\eta}^{(2)}_{j_1, j_2} = \begin{cases}
\frac{C^{(2)}_{j_1, j_2}}{\hat{S}_{j_1, j_2}}, & |j_1| + |j_2| = 2,\\
0, &|j_1| + |j_2| \neq 2.
\end{cases}
\end{equation}

One can obtain the asymptotic expansions of quasi-periodic traveling waves in the case of deep water by letting 
$h$ go to infinity. In this case, the expressions of $\td \eta^{(1)}$, $b^{(0)}$ and $\tau^{(0)}$ read
\begin{equation}
\tilde{\eta}^{(1)} = \hat{\eta}_{1,0}e^{i\alpha_1} + \hat{\eta}_{0,1}e^{i\alpha_2} + c.c., \qquad
 b^{(0)} = g+ \frac{g}{k}, \qquad \tau^{(0)} = \frac{g}{k}
\end{equation}
and the expressions of $\td \eta^{(2)}$, $b^{(1)}$ and $\tau^{(1)}$ read
\begin{equation*}
\td \eta^{(2)} = \hat{\eta}_{2,0}^{(2)}e^{i(2\alpha_1)} + \hat{\eta}_{0,2}^{(2)}e^{i(2\alpha_2)}
+ \hat{\eta}_{1,1}^{(2)}e^{i(\alpha_1 + \alpha_2)} + \hat{\eta}_{1,-1}^{(2)}e^{i(\alpha_1-\alpha_2)}
+ c.c., \qquad b^{(1)} = \tau^{(1)} = 0,
\end{equation*}
\begin{align} \label{eq:eta:order:2}
\hat{\eta}_{2,0}^{(2)} &= -g\hat{\eta}_{1,0}^2\frac{(2k-1)/k}{\hat S_{2,0}}, \qquad \qquad 
&\hat{\eta}_{0,2}^{(2)} &= g\hat{\eta}_{0,1}^2\frac{k(k-2)}{\hat{S}_{0,2}},\\ 
\hat{\eta}_{1,1}^{(2)} &= -g\hat{\eta}_{1,0}\hat{\eta}_{0,1}\frac{(k+1)}{\hat{S}_{1,1}}, \qquad  \qquad 
&\hat{\eta}_{1,-1}^{(2)} &= -g\hat{\eta}_{1,0}\hat{\eta}_{0,1}\frac{(k+1)}{\hat{S}_{1,-1}}, \notag
\end{align}
where 
\begin{equation}\label{eq:S:deep}
\hat{S}_{j_1, j_2} = \frac{g}{k}(|j_1+kj_2| - k)(|j_1 + k j_2|-1).
\end{equation}

Even though we stop at the second order in the weakly nonlinear model, one can continue computing higher-order terms by induction. 
Suppose that we have obtained terms of order $n-1$ for $\tilde \eta$ and terms of order
$n-2$ for $b$ and $\tau$.
Eliminating the coefficients of $\epsilon^n$ in (\ref{eq:trav:torus:conf}),  
we find that 
\begin{equation}
g\tilde \eta^{(n)} - b^{(0)}H^{\coth}\big[\tilde\eta^{(n)}_\alpha\big] 
-\tau^{(0)}\tilde\eta^{(n)}_{\alpha\alpha}
- b^{(n-1)}H^{\coth}\big[\tilde\eta^{(1)}_\alpha\big]
- \tau^{(n-1)} \tilde \eta^{(1)}_{\alpha\alpha} = C^\e{n},
\end{equation}
where $C^\e{n}$ depends on $\left\{b^{(j)}\right\}_{0\leq j\leq n-2}$,
$\left\{\tau^{(j)}\right\}_{0\leq j\leq n-2}$ and
$\left\{\tilde\eta^{(j)}\right\}_{0\leq j\leq n-1}$.  Comparing
  the Fourier coefficients of both sides of the above equation, we
  have
\begin{equation}
\hat{S}_{j_1, j_2} \hat{\eta}^{(n)}_{j_1, j_2} 
+ \hat{Q}_{j_1, j_2}^{(n-1)}\hat{\eta}^{(1)}_{j_1, j_2}
= \hat{C}^{(n)}_{j_1, j_2},
\end{equation}
where $\hat S_{j_1, j_2}$ and $\hat Q_{j_1, j_2}^{(n-1)}$ are given in \eqref{eq:S} and \eqref{eq:B:2}, respectively.
Eventually  we can express $b^{(n-1)}$, $\tau^{(n-1)}$ and the Fourier coefficients of $\td \eta^{(n)}$  as follows,
\begin{equation}\label{eq:asym:induction}
\begin{gathered}
\hat{\eta}^{(n)}_{j_1, j_2} = \frac{\hat{C}^{(n)}_{j_1, j_2}}{\hat{S}_{j_1, j_2}}, \qquad\qquad 
 (j_1, j_2) \neq (\pm1, 0),\, (0, \pm1), \\ 
b^{(n-1)} = 
\frac{\frac{\hat{C}_{0,1}^{(n)}}{\hat{\eta}_{0,1}} - k^2 \frac{\hat{C}_{1,0}^{(n)}}{\hat{\eta}_{1,0}} }
{k(k\coth(h) - \coth(kh))},
\qquad \qquad \tau^{(n-1)} = 
\frac{\coth(h)\frac{\hat{C}_{0,1}^{(n)}}{\hat{\eta}_{0,1}}- k\coth(kh)\frac{\hat{C}_{1,0}^{(n)}}{\hat{\eta}_{1,0}} }{k(k\coth(h) - \coth(kh))}.
\end{gathered}
\end{equation}
Note that the Fourier coefficients of $\tilde \eta^{(n)}$ are obtained
through a division by $\hat{S}_{j_1, j_2}$ for $(j_1, j_2) \neq (\pm1,
  0), (0, \pm1)$.  If the $\hat{S}_{j_1, j_2}$ can become arbitrarily
small, the corresponding terms $\hat{\eta}^{(n)}_{j_1, j_2}$ may be
strongly amplified, calling into question the nature of the expansion
\eqref{eq:asym:expansion}. This is known as a small divisor problem.
In the case of deep water, it is clear from \eqref{eq:S:deep} that
some of the $\hat{S}_{j_1, j_2}$ approach zero as $|j_1|, |j_2|$ grow
without bound. Speculating on the possibilities, it may be that
\eqref{eq:asym:expansion} becomes an asymptotic series provided that $k$
is sufficiently irrational, satisfying a diophantine condition
\cite{moser1966theory}
\begin{equation}\label{eq:diophantine}
  |k - j_1/j_2| > C|j_2|^{-\nu}, \qquad \qquad
  j_1\in \mbb{Z}, \, j_2 \in \mbb{Z}\backslash \{0\},
\end{equation}
where $C$ is a positive constant and $\nu> 2$. But it may also be that
exact mathematical solutions only exist for sufficiently small values
of $\veps$ in a totally disconnected Cantor-like set \cite{iooss05},
even under the assumption \eqref{eq:diophantine}. More research is
needed to resolve these questions.

The story is even more complicated in the case where the fluid is of
finite depth because the expression for $\hat{S}_{j_1, j_2}$ involves
the hyperbolic cotangent function. But this formula becomes simpler
again in the case of shallow water, where $h$ is small. Expanding
$\coth(h)$ and $\coth(kh)$ in \eqref{eq:S} in a Laurent expansion
about $h=0$, we obtain
\begin{equation}\label{eq:S:asym:h}
\hat{S}_{j_1, j_2} = \frac{gh^4}{45}(|j_1+kj_2|^2-k^2)(|j_1+kj_2|^2-1) + O(h^6).
\end{equation}
We notice that $\hat{S}_{j_1, j_2}$ can be very small due to the
factor of $h^4$ in \eqref{eq:S:asym:h}.  Thus, in the shallow water
regime, the amplitudes of quasi-periodic traveling waves bifurcating
from the zero-amplitude solution must be small, with $\veps$ at most
$O(h^4)$, if weakly nonlinear theory is to predict their behavior.



\section{Numerical methods and results} \label{sec:numerical}

As in Section~\ref{sec:traveling} above,
we focus our discussion on quasi-periodic functions with two
quasi-periods.  All computation will be performed with respect to
torus functions on $\mbb T^2$; the one-dimensional quasi-periodic
functions will be reconstructed from the torus functions using
(\ref{eq:general_quasi_form}).  Let $f(\alpha)$ be a quasi-periodic
function with two quasi-periods and let $\tilde f$ denote the
corresponding periodic function on $\mbb T^2$,
\begin{equation}\label{quasi_f_form}
  f(\alpha) = \tilde f(\alpha,k\alpha), \qquad
  \tilde f(\alpha_1,\alpha_2) = 
  \sum\limits_{j_1, j_2 \in \mathbb{Z}}
  \hat{f}_{j_1, j_2} e^{i(j_1\alpha_1+j_2\alpha_2)}, \qquad
  (\alpha_1, \alpha_2)\in\mathbb{T}^2.
\end{equation}
Following \cite{quasi:ivp, quasi:trav}, we adopt a pseudo-spectral
method and represent $\tilde f$ in two ways:

\begin{itemize}
\item[(1)] Via the values of $\td f$ on a uniform $M_1\times M_2$ grid on the torus $\mbb T^2$,
\begin{equation}
  \tilde f_{m_1,m_2} = \tilde f(2\pi m_1/M_1\,,\,2\pi m_2/M_2), \qquad
  0\le m_1<M_1\,,\,0\le m_2<M_2.
\end{equation}

\item[(2)] Via the truncated two-dimensional Fourier series of $\td f$, with Fourier coefficients given by
\begin{equation}
\hat f_{j_1,j_2} = \frac1{M_2}\sum_{m_2=0}^{M_2-1}
  \left(\frac1{M_1}\sum_{m_1=0}^{M_1-1}
    \tilde f_{m_1,m_2} e^{-2\pi ij_1m_1/M_1}\right)e^{-2\pi ij_2m_2/M_2}, 
\end{equation}
where $ 0\le j_1\le M_1/2, -M_2/2< j_2\le M_2/2.$
\end{itemize}
We use the `r2c' and 'c2r' version of the 2d FFTW library to rapidly
transform between these two forms.
Products, powers and quotients in (\ref{eq:time:evol:torus})
and (\ref{eq:trav:torus:conf}) are evaluated point-wise on the grid 
while derivatives and Hilbert transforms are computed in Fourier space via Definition~\ref{def:deriv}
and \ref{def:hilbert}. 
In the scope of this paper, we choose $k = 1/\sqrt{2}$ for all numerical examples.


\subsection{Time evolution of spatially quasi-periodic waves of finite depth}
\label{sec:ivp}
To compute the time evolution of spatially quasi-periodic waves, we discretize~(\ref{eq:time:evol:torus}) on $\mbb T^2$ and use the fifth-order explicit Runge-Kutta method of Dormand and Prince
\cite{hairer:I, quasi:ivp}.
The initial condition of the water wave is given in physical space,
which is more natural in practice, and we compute the conformal
mapping to transform the initial condition to conformal space using
the method described in Appendix \ref{sec:conformal:transform}. The
numerical examples discussed below are gravity waves but our numerical
method also applies to the case of nonzero surface tension.

\begin{figure}[h]
\begin{centering}
\includegraphics[width=\textwidth]{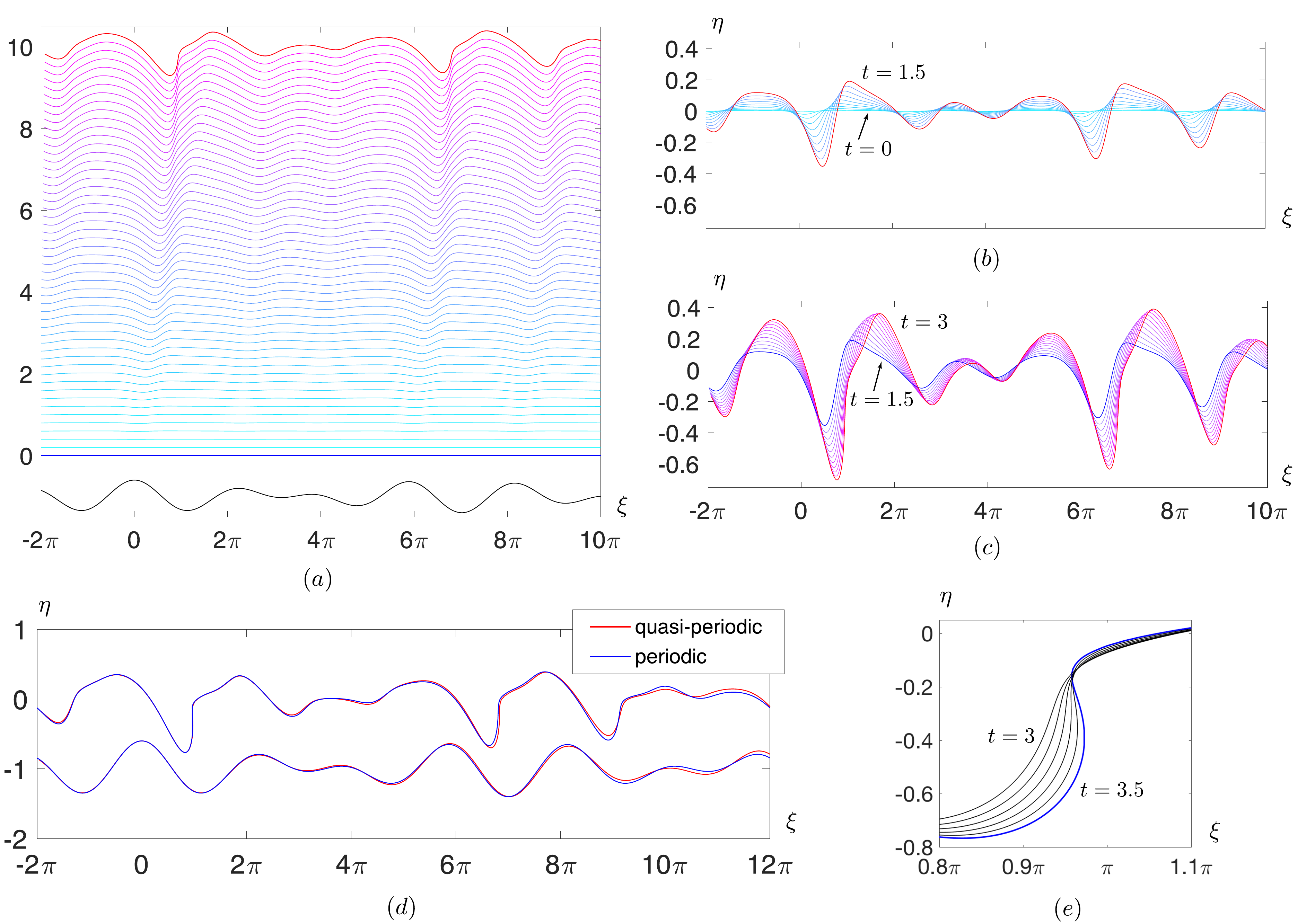}
\caption{\label{fig:background} Panels (a)--(c) show the time
  evolution from $t = 0$ to $t = 3$ of an initially flat free
  surface in the presence of a background flow and a quasi-periodic
  bottom boundary.  In panel (a), the bottom boundary is plotted
    in black and we added an upward spatial shift, given by
    \eqref{eq:upshift:kdv}, to separate the curves from each other.
    Panels (b) and (c) show details of the free surface without such a
    shift.  In panel (d), we evolve this solution further, to $t=3.5$,
    and compare it to a periodic calculation with a bottom boundary of
    similar shape over the spatial range shown. Both waves overturn
    near $\xi = \pi$ and $t=3.3$. The evolution of the overturning
    periodic wave from $t=3$ to $t=3.5$ is shown in panel (e).  }
\end{centering}
\end{figure}

Figure \ref{fig:background} shows the time evolution of a free surface
wave that is initially flat and develops quasi-periodic dynamics in
the presence of a background flow and a quasi-periodic bottom
boundary. In physical space, the bottom boundary is parameterized by
\begin{equation}
\eta^{\btm, \ph}(x) = -1 + 0.2\cos(x) + 0.2\cos(x/\sqrt{2})
\end{equation}
 and the mean velocity of the background flow in (\ref{eq:varphi:sfc})
 is $\mc U = 1$.  In the computation, we use $M_1 = M_2 = 512$ and
 compute the time evolution of the wave from $t= 0$ to $t=3$ with time
 steps $\Delta t = 10^{-5}$.  In panel (a), the black line plots the
 bottom boundary and the blue line plots the flat free surface at $t =
 0$.  To better distinguish the shape of the free surface at different
 times, we add an upward spatial shift to each
   curve. The time difference between two adjacent curves is 0.06,
 and we plot
\begin{equation}\label{eq:upshift:kdv}
\eta^\sfc(\alpha, t_n) + 10t_n/3, \qquad t_n = 0.06n, \quad n = 0, 1, \ldots, 50.
\end{equation}
Due to the background flow and the quasi-periodic bottom boundary, the
free surface wave moves from left to right and forms wave crests ahead
of the peaks of the bottom boundary, which deflects the fluid upward.
Panels (b) and (c) show snapshots of the time evolution of the free
surface from $t = 0$ to $t = 1.5$ and from $t=1.5$ to $t = 3$
separately without the upward shift given in (\ref{eq:upshift:kdv});
the time difference between two adjacent curves in both panels is
0.15.

In panel (d) of Figure~\ref{fig:background}, we further evolve the
quasi-periodic wave from $t = 3$ to $t = 3.5$ using the same
$512\times512$ spatial grid on the torus and plot the free surface at
$t = 3.5$ with a red line. We find that the wave overturns near $\xi =
\pi$ at $t=3.308$. However, the quasi-periodic wave is under-resolved
by $t=3.5$, with gridpoints spread out enough to see discrete line
segments in the plot near $\xi=\pi$ and small oscillations visible
near $\xi=\pi$ and $\xi = 7\pi$.  In the infinite depth case
\cite{quasi:ivp}, refining the grid to $4096\times4096$ was sufficient
to resolve an overturning spatially quasi-periodic wave so that the
Fourier modes decay to $10^{-12}$ at all times. Here, rather than
refine the mesh beyond $512\times512$, we compute the time evolution
of a periodic wave under the same initial condition and background
flow, but with a periodic bottom boundary
\begin{equation}
\eta^{\btm, \ph}(x) = -1 + 0.2\cos(x) + 0.2\cos(5x/7),
\end{equation}
where $5/7$ is a rational approximation of $k = 1/\sqrt{2}$.  We use
the spatial resolution $M = 32768$ for the periodic calculation (which
  employs 16385 Fourier modes). The free surface of the periodic wave
at $t = 3.5$ is plotted with a blue line in panel (d).  We observe
that the two waves in panel (d) resemble each other, but are not
exactly the same, due to the different bottom boundaries. They will
differ much more for larger values of $|x|$, where $\cos(x/\sqrt2)$ is
farther from $\cos(5x/7)$. The periodic wave overturns near $\xi =
\pi$ at time $t=3.288$. We show the time evolution of the periodic
wave near the overturning point from $t = 3$ to $t = 3.5$ in panel
(e), where the time difference between adjacent curves is
$\Delta t = 0.1$.

\begin{figure}[t]
\begin{centering}
\includegraphics[width=\textwidth]{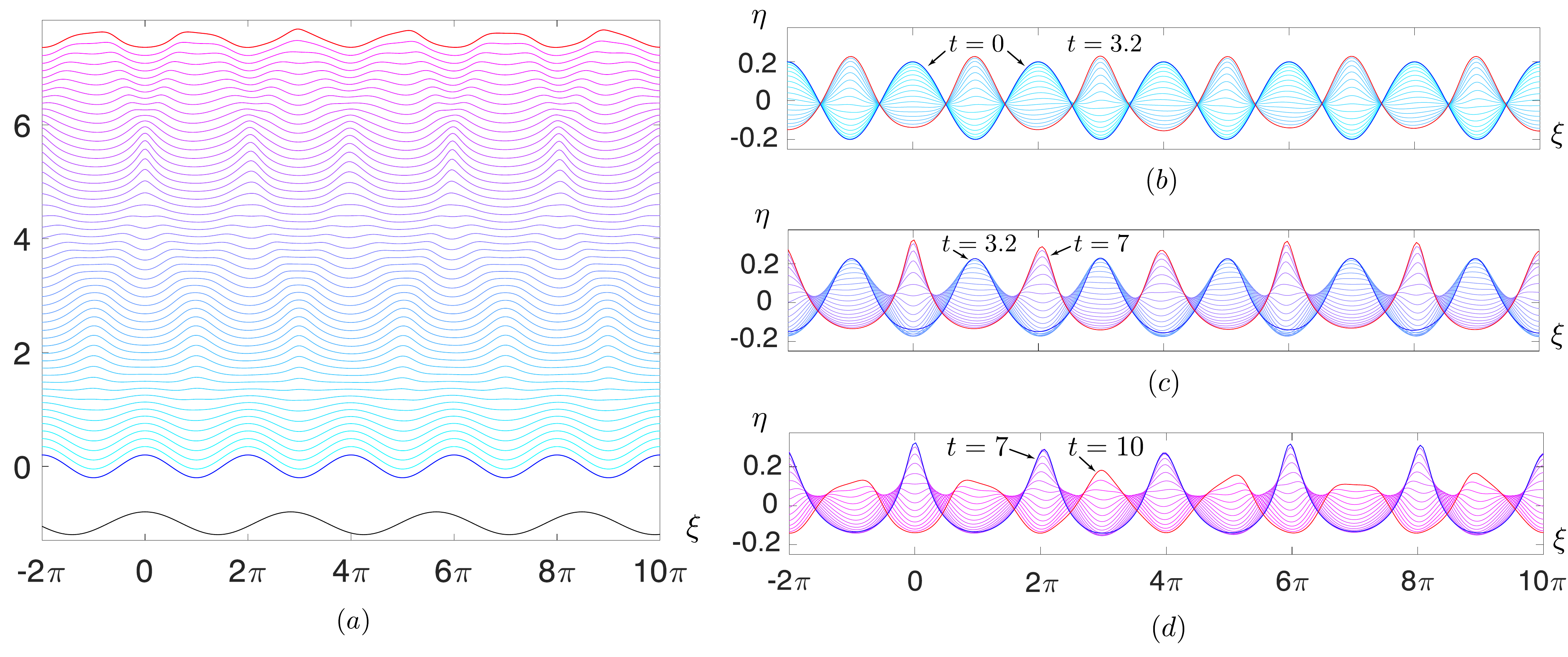}
\caption{\label{fig:standing} Time evolution from $t = 0$ to $t
    = 10$ of an initially periodic free surface evolving from
    rest, with no background flow, over a periodic bottom boundary
  whose spatial period is irrationally related to the initial period
  of the free surface.  Panel (a) shows the free surface at
    different times with the upward shift \eqref{eq:upshift:wilton};
    the bottom boundary is plotted in black. Panels (b), (c) and (d)
    show details of the free surface evolution without the shift.}
\end{centering}
\end{figure}

Figure \ref{fig:standing} shows the time evolution of an initially periodic free surface wave in the presence of a periodic bottom boundary whose spatial period is irrationally related to the initial condition. 
In physical space, the initial free surface and the bottom boundary are given by
\begin{equation}
\eta^{\sfc, \ph}_0(x) = 0.2\cos(x), \qquad \qquad 
\eta^{\btm, \ph}(x) = -1+0.2\cos(x/\sqrt{2}).
\end{equation}
In panel (a), the initial free surface and the bottom boundary are plotted with blue and black curves, respectively. As shown in the figure, 
they are both periodic and the bottom boundary's wavelength is longer than that of the free surface. 
We use  $M_1 = M_2 = 256$ in the computation and  evolve the water wave from $t = 0$ to $t = 10$ with time steps $\Delta t = 2\times 10^{-5}$. At $t=0$, the fluid is at rest with zero velocity potential. 
In panel (a), we plot the time evolution of the  free surface with an upward spatial shift
\begin{equation}\label{eq:upshift:wilton}
\eta^\sfc(\alpha, t_n) + 0.75t_n, \qquad t_n = 0.2n, \quad n = 0, 1, \ldots, 50. 
\end{equation}
The free surface flattens due to the force of gravity and rises again
due to inertia, which is similar to the oscillation of a standing
water wave \cite{mercer:92, water1}.  One can observe that the crests
and troughs of the surface wave are not symmetric for $t>0$ except at
$x = 0$ due to the even symmetry of the initial condition.  In panels
(b), (c) and (d), we plot snapshots of the time evolution of the free
surface without the upward shift (\ref{eq:upshift:wilton}) from $t = 0$
to $t = 3.2$; from $t = 3.2$ to $t = 7$; and from $t = 7$ to $t =
10$. The time difference between two adjacent curves is 0.2. One can
observe that the wave oscillates up and down like a standing
wave. However, as a consequence of the quasi-periodic interactions
between the surface wave and the bottom boundary, the heights of
different crests are different at any given time.


\subsection{Spatially quasi-periodic traveling waves}
\label{sec:num:trav}


We formulate the traveling wave problem as a nonlinear least-squares
problem, which we solve using a variant of the Levenberg-Marquardt
algorithm~\cite{wilkening2012overdetermined, nocedal, quasi:trav}.  In
Section \ref{sec:govern:eq:quasi:trav}, we introduced the residual
function $\mc{R}$ in (\ref{eq:trav:torus:conf}), which depends on
$\tau$, $b$, $\td \eta$, and demonstrated that the solutions of the
traveling wave problem are the solutions of $\mc{R}[\tau, b, \td \eta]
= 0$.  In the computation, we consider $\tau$, $b$ and the Fourier
coefficients of $\td \eta$ as unknowns, denoted $\hat \eta$, and
define the following scalar objective function
\begin{equation}\label{eq:FR:def}
  \mathcal{F} [\tau, b, \hat\eta] := \frac{1}{8\pi^2}\int_{\mathbb{T}^2}
  \mathcal{R}^2[\tau, b, \hat\eta]\,\, d\alpha_1\,d\alpha_2.
\end{equation}
Note that solving (\ref{eq:trav:torus:conf}) is equivalent to finding a zero
of the objective function $\mathcal{F}[\tau, b, \hat\eta]$. 
For the unknown $\hat{\eta}$, we only vary the leading Fourier coefficients 
$\hat{\eta}_{j_1, j_2}$ with $|j_1|\leq N_1< M_1/2$, $|j_2|\leq N_2 < M_2/2$ and set the other Fourier coefficients to zero.
According to the assumption~(\ref{eq:eta:assump}), we also set $\hat{\eta}_{0,0} = 0$ and
require that the Fourier coefficients $\hat \eta_{j_1, j_2}$ are real and satisfy $\hat \eta_{-j_1, -j_2} = \hat \eta_{j_1, j_2}$ . 
Consequently, the number of independent leading Fourier coefficients is
 $N_\text{tot} = N_1(2N_2+1)+N_2$.
As discussed in Section \ref{sec:govern:eq:quasi:trav}, we choose $\hat{\eta}_{1,0}$, $\hat{\eta}_{0,1}$ and $h$ as bifurcation parameters when computing quasi-periodic traveling solutions
bifurcating from the zero-amplitude solution 
and fix them at nonzero amplitudes in the minimization of $\mc F$. 
Therefore there are $N_\text{tot}$ parameters to compute, which are stored in a vector $\bds p$ as follows
\begin{equation}\label{eq:p:enum}
  p_1=\tau, \quad p_2=\hat\eta_{1,1}, \quad
  p_3=b, \quad p_4=\hat\eta_{1,-1}, \quad
  p_5=\hat\eta_{0,2}\;\;,\;\; \dots\;\;,\;\;
  p_{N_\text{tot}}=\hat\eta_{1,-N_2}.
\end{equation} 
The Fourier modes have been organized in a spiral fashion so that low frequency modes appear first in the list and $\hat{\eta}_{1,0}$, $\hat{\eta}_{0,1}$ have been replaced by $\tau$ and $b$; see \cite{quasi:trav} for details. Our goal is to find $\bds p$ given $\hat\eta_{1,0}$ and $\hat\eta_{0,1}$ such that $\mc
F[\bds p;\hat\eta_{1,0},\hat\eta_{0,1}]=0$, where we have re-ordered the arguments of
$\mc F$ and $\mc R$ in \eqref{eq:FR:def}.
In the computation, the function $\mc{R}$ is evaluated at $M_1\times M_2$ grid points, hence there are $M_1 M_2$ equations, which are more than the number of unknowns. 
For this reason, the nonlinear least-squares problem is overdetermined.

The objective function $\mc{F}$ is computed from
$\mc{R}$  by the trapezoidal rule
approximation over $\mbb T^2$, which is spectrally accurate,
\begin{equation} \label{eq:numerical_objective}
  \begin{aligned}
    f(\bds p) &= \frac{1}{2} r(\bds p)^Tr(\bds p) \approx
    \mathcal{F}\left[\bds p;\hat\eta_{1,0},\hat\eta_{0,1} \right], \\ 
    r_m(\bds p) &= \frac{\mathcal{R}\left[\bds p;\hat\eta_{1,0},\hat\eta_{0,1}\right]
      (\alpha_{m_1}, \alpha_{m_2})}{\sqrt{M_1M_2}},
  \end{aligned}
    \quad \left(\begin{gathered}
      m = 1+m_1+M_1m_2 \\
      \alpha_{m_i} = 2\pi m_i/M_i
    \end{gathered}\right),
       \quad 0\leq m_i < M_i.
\end{equation}
The parameters $p_j$ are chosen to minimize $f(\bds p)$ using the
Levenberg-Marquardt method~\cite{wilkening2012overdetermined,nocedal}. 
The method requires a Jacobian matrix $\partial r_m/\partial p_j$,
which we compute by solving the variational equations (\ref{eq:variational:eq}).
We have $\der{r_m}{p_j} =
\delta\mc R(\alpha_{m_1},\alpha_{m_2})/\sqrt{M_1M_2}$,
where $m=1+m_1+M_1m_2$ and the
$j$th column of the Jacobian corresponds to setting $\delta p_j$ in (\ref{eq:p:enum}) to 1 and the others to 0 depending on the perturbation direction:
$\delta\tau$, $\delta b$ or $\delta\hat\eta_{j_1,j_2}$.

\begin{figure}[t]
\begin{centering}
\includegraphics[width=0.8\textwidth]{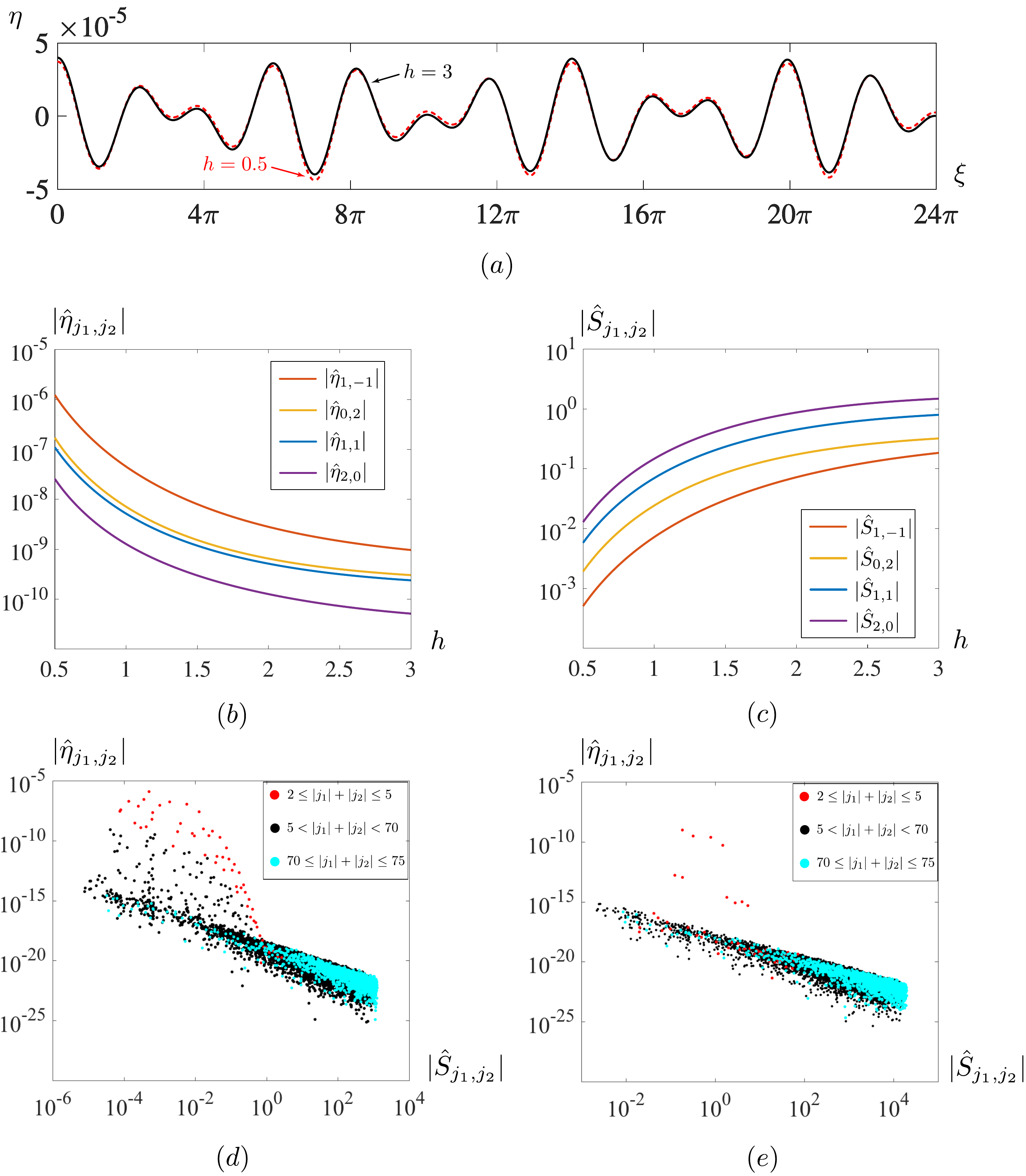}
\caption{\label{fig:small_amplitude} Quasi-periodic traveling
  gravity-capillary waves bifurcating from the zero solution.
  (a) Surface elevation function of two solutions with
  $h=0.5$ (dashed red line) and $h=3.0$ (solid black line). (b)
  Amplitudes of Fourier coefficients $\hat{\eta}_{2,0}$,
  $\hat{\eta}_{0,2}$, $\hat{\eta}_{1,1}$, $\hat{\eta}_{1,-1}$ of
  quasi-periodic traveling solutions for which $\hat{\eta}_{1,0}$ and
  $\hat{\eta}_{0,1}$ are fixed at $10^{-5}$. (c)
  Absolute value of the corresponding divisors $\hat{S}_{j_1,j_2}$
  defined by (\ref{eq:S}) in the weakly nonlinear model
  (\ref{eq:eta:order:2}). Here we solve \eqref{eq:trav:torus:conf}
  by minimizing $f(\bds p)$
  in \eqref{eq:numerical_objective} and
  check whether the solution behaves as predicted by
  (\ref{eq:eta:order:2}). Panels (d) and (e) show
  $|\hat{\eta}_{j_1,j_2}|$ versus $|\hat{S}_{j_1,j_2}|$ for $2\leq
  |j_1| +|j_2|\leq 75$ in the cases where $h = 0.5$ and $h = 3$,
  respectively.  }
\end{centering}
\end{figure}

We compute quasi-periodic traveling solutions that bifurcate from the
zero solution using $N_x = N_y = 75$ and $M_x = M_y = 200$. We fix
$\hat{\eta}_{1,0} = \hat{\eta}_{0,1} = 10^{-5}$, choose $h$ to be the
continuation parameter, and decrease $h$ from 3 to 0.5 with $\Delta h
= 0.01$ to obtain a family of quasi-periodic solutions.  In panel (a)
of Figure \ref{fig:small_amplitude}, we plot the wave profile of the
free surface for solutions at $h = 0.5$ and $h = 3$. The difference
between these two solutions is small because they are both
small-amplitude bifurcations from the zero solution for which we
imposed the same amplitude parameters $\hat{\eta}_{1,0}$ and
$\hat{\eta}_{0,1}$ at linear order.  We stayed close to the linear
regime in this example to investigate whether traveling solutions of
the fully nonlinear equations, which we compute using the
Levenberg-Marquardt method, behave as predicted by weakly nonlinear
theory. While the wave profiles are close to one another, the values
of $h$ (3 and 0.5) and $\tau$ (1.23088845108 and 0.0812490184995)
differ substantially for the two solutions.

In panel (b) of Figure \ref{fig:small_amplitude}, we plot the absolute
value of the leading Fourier coefficients $|\hat{\eta}_{2,0}|$,
$|\hat{\eta}_{0,2}|$, $|\hat{\eta}_{1,1}|$ and $|\hat{\eta}_{1,-1}|$
of the computed solutions as functions of $h$, holding
$\hat{\eta}_{1,0}$ and $\hat{\eta}_{0,1}$ fixed at $10^{-5}$. These
Fourier coefficients decrease as $h$ increases. In panel (c), we plot
the absolute value of the divisors $\hat{S}_{j_1,j_2}$ defined in
(\ref{eq:S}) corresponding to these four Fourier coefficients, which
decrease as $h$ decreases. The behavior of the Fourier coefficients
and $\hat{S}_{j_1, j_2}$ is consistent with the weakly nonlinear
approximations (\ref{eq:eta:order:2}), where the Fourier coefficients
are obtained through division by $\hat{S}_{j_1, j_2}$. As a result,
smaller values of $\hat{S}_{j_1, j_2}$ lead to larger Fourier
coefficients. Note that we are checking whether traveling solutions of
the Euler equations \eqref{eq:trav:torus:conf} obtained by minimizing
$f(\bds p)\approx \mathcal{F}\left[\bds
  p;\hat\eta_{1,0},\hat\eta_{0,1} \right]$ in
\eqref{eq:numerical_objective} via the Levenberg-Marquardt method
behave as predicted by the weakly nonlinear model
\eqref{eq:eta:order:2}; we did not solve \eqref{eq:eta:order:2}
directly.

Panels (d) and (e) of Figure \ref{fig:small_amplitude} demonstrate the
relationship between $|\hat{\eta}_{j_1, j_2}|$ and $|\hat S_{j_1,
  j_2}|$ with $2\leq |j_1| + |j_2| \leq 75$ for traveling solutions at
$h = 0.5$ and $h = 3$, respectively.  Since the largest Fourier
coefficients are fixed at $10^{-5}$, one expects roundoff errors
around $10^{-20}$. But instead the ``roundoff floor,'' visible in both
panels, appears to grow linearly as $|\hat S_{j_1, j_2}|$
decreases. This suggests that roundoff errors in the
Levenberg-Marquardt method are amplified by the reciprocals of the
divisors $\hat{S}_{j_1, j_2}$ even though this is not a weakly
nonlinear calculation. The ``active'' modes in which
$\big|\hat\eta_{j_1,j_2}\big|$ extends above the roundoff floor appear
to be well-resolved. The plots look nearly identical if we refine the
calculation, keeping $N_x = N_y = 75$ but increasing $M_x$ and $M_y$
from 200 to 300. In fact, we plotted the data from this finer mesh in panels
(d) and (e). In panel (e), when $h=3$, there are just a few active
modes $\hat\eta_{j_1,j_2}$, and they all correspond to low frequency
modes with $2\le |j_1|+|j_2|\le 5$. But in panel (d), when $h=0.5$,
there are many active modes of both small and intermediate frequency,
plotted with red and black markers, respectively. This is consistent
with \eqref{eq:S:asym:h} and panel (c), where the small divisors from
weakly nonlinear theory decrease as $h$ decreases. The fixed values
$\hat{\eta}_{1,0} = \hat{\eta}_{0,1} = 10^{-5}$ we selected for this
calculation appear to be small enough when $h=3$ that we could have
computed the solution by weakly nonlinear theory, but large enough at
$h=0.5$ that it was necessary to solve the problem by the
Levenberg-Marquardt approach.

Next we search for quasi-periodic bifurcations from finite-amplitude
periodic traveling waves of finite depth. We use a new procedure,
described in detail for the case of deep water in \cite{quasi:bifur},
to locate bifurcation points. Specifically, we use the signed smallest
singular value $\chi(s)$ of the Jacobian $\mc J^\qua(s)$, as in
\eqref{eq:chi:def}, as a bifurcation ``test function'' that changes
sign at bifurcation points. When a zero of $\chi(s)$ is found, the
kernel of the Jacobian $\mc J^\qua(s)$ of \eqref{eq:chi:def} also
furnishes a search direction $\delta\td\eta^\qua$ for the
quasi-periodic branch. We use $\td \eta^\per + \epsilon
\delta\td\eta^\qua$ with an empirically chosen value of $\veps$ as the
initial guess for the Levenberg-Marquardt solver. We then use
numerical continuation to follow this branch beyond the realm of
linearization about the periodic traveling wave.  Instead of using
$\hat\eta_{1,0}$, $\hat\eta_{0,1}$ and $h$ as continuation parameters,
we use $\tau$, $h$ and the Fourier mode $\hat\eta_{0,1}$.  For
simplicity, we hold $\tau$ and $h$ fixed and just vary the Fourier
mode to obtain a one-parameter family of quasi-periodic solutions.

\begin{figure}[b]
\begin{centering}
\includegraphics[width=0.7\textwidth]{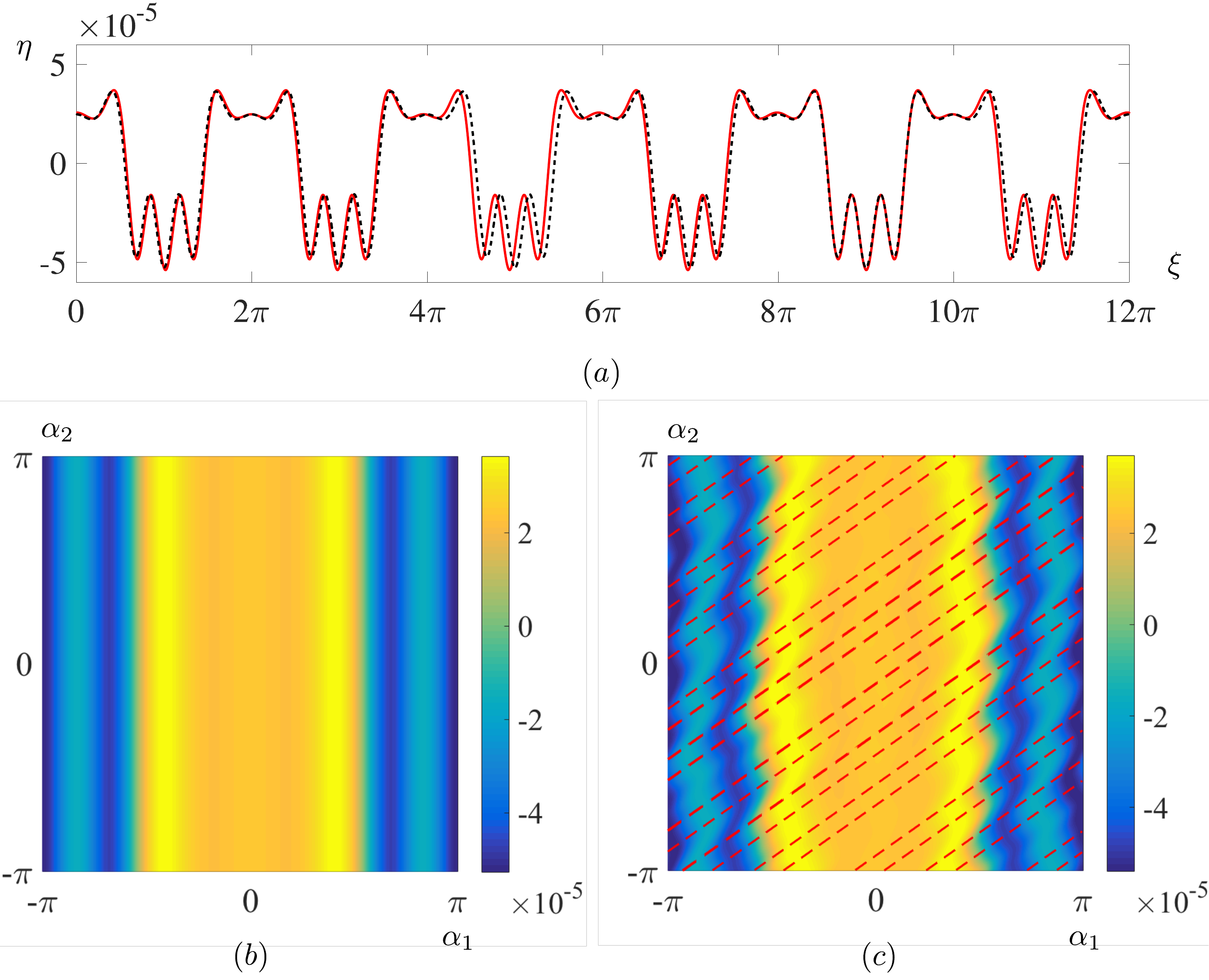}
\caption{\label{fig:wilton_small} Quasi-periodic bifurcation from a periodic
  traveling gravity-capillary wave.  
  Panel (a) shows the periodic traveling wave where a bifurcation was found
  and the largest-amplitude solution we computed on the quasi-periodic bifurcation
  branch. The dotted black line corresponds to the periodic wave and
  the red line corresponds to the quasi-periodic wave.
  Panels (b) and (c) show contour plots of the torus functions of the periodic wave
  and the quasi-periodic wave, respectively. 
  The 1D quasi-periodic wave in panel (a) is extracted from the corresponding 
  torus function along the characteristic lines of slope $k=1/\sqrt2$, plotted
  with red dashed lines in panel~(c).}
\end{centering}
\end{figure}

\begin{figure}[t]
\begin{centering}
\includegraphics[width=0.7\textwidth]{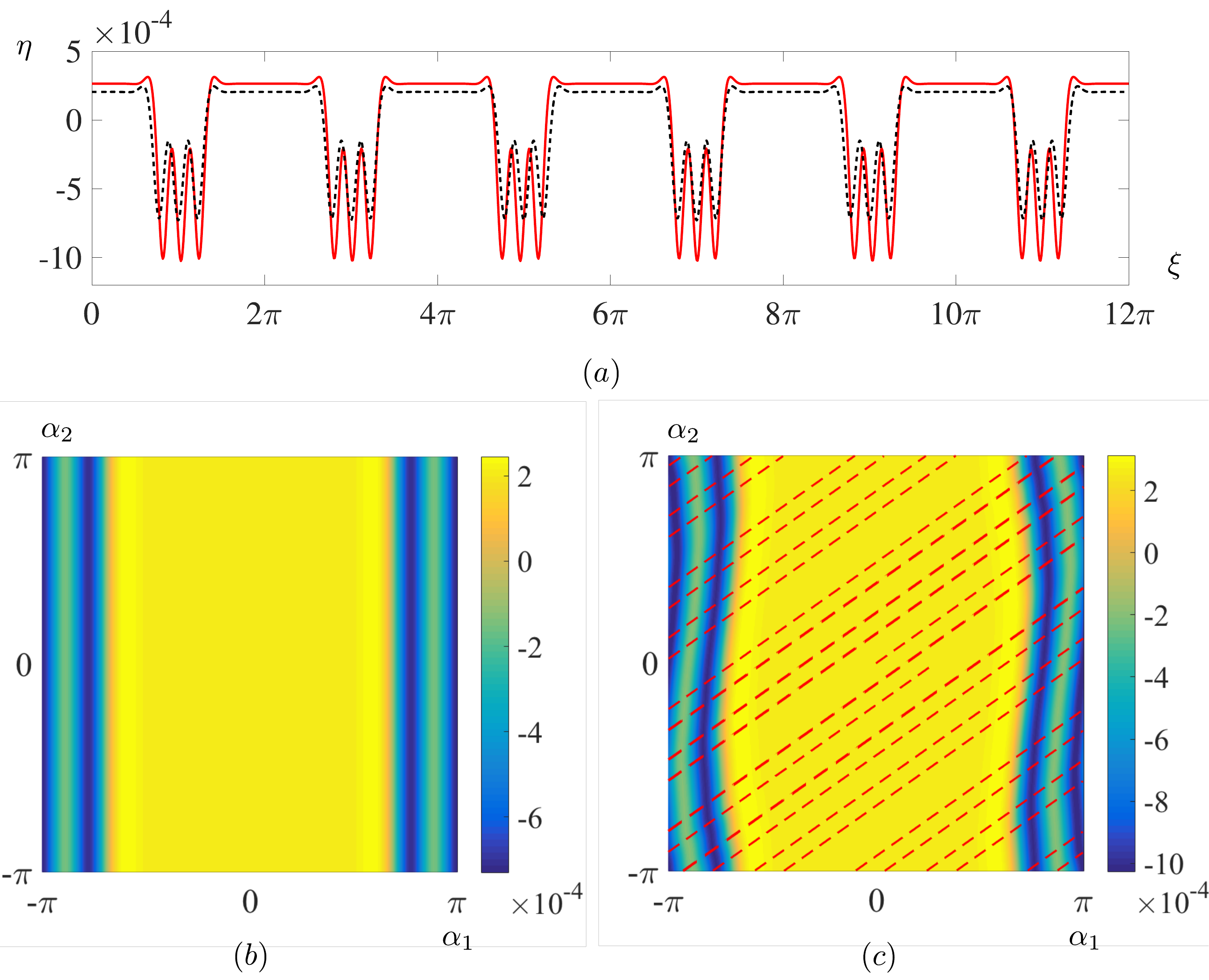}
\caption{\label{fig:wilton_large} Quasi-periodic bifurcation from a
  larger-amplitude periodic traveling gravity-capillary wave.  The
  panels show the same information as in
  Figure~\ref{fig:wilton_small}.}
\end{centering}
\end{figure}

Figures \ref{fig:wilton_small} and \ref{fig:wilton_large} show two quasi-periodic
gravity-capillary waves bifurcating from a branch of periodic traveling waves. 
The fluid depth in conformal space is $h = 0.1$. 
We set $\tau =  0.00327672209262$ so that the first Fourier mode of the periodic waves resonates with the fifth Fourier mode, which corresponds to solutions of the Wilton ripple problem
\cite{akers2012wilton, trichtchenko:16, akers2021wilton}. 
For the
periodic traveling wave, we set $M = 300$, $N = 100$ and use
$s=\hat{\eta}_{1}$ as the continuation parameter. 
The 1D waves are
computed on $\mbb T$ and embedded in $\mbb T^2$ when searching for
bifurcations, so that $\hat\eta_1$ becomes $\hat\eta_{1,0}$. We
computed periodic waves with amplitude $s$ ranging from $10^{-5}$ to
$2\times 10^{-4}$ with $\Delta s = 10^{-5}$. By tracking the sign of $\chi(s)$, we find out that there is a zero of $\chi(s)$
when $s$ belongs to intervals 
$[10^{-5}, 2\times 10^{-5}]$, $[4\times 10^{-5}, 5\times 10^{-5}]$, $[7\times 10^{-5}, 8\times 10^{-5}]$,
$[1.1\times 10^{-4}, 1.2\times 10^{-4}]$ and $[1.7\times 10^{-4}, 1.8\times 10^{-4}]$. We focus our discussion on the first and last intervals and locate the zeros of $\chi(s)$ in these intervals, which are the bifurcation points, using the numerical algorithm described in \cite{quasi:bifur}. 
In double precision, the zeros and corresponding values of $\chi$ are
\begin{equation}
\begin{aligned}
s_1 &= 1.83810709940\times 10^{-5}, &\quad \chi(s_1) &= -7.8 \times 10^{-15}, \\
s_2 &= 1.72625902886\times 10^{-4}, &\quad \chi(s_2) &= 4.8\times 10^{-15}.
\end{aligned}
\end{equation}
The periodic solutions at $s_1$ and $s_2$ are plotted with dotted
black lines in panel~(a) of Figures \ref{fig:wilton_small} and
\ref{fig:wilton_large}, respectively.  These periodic solutions
demonstrate the nonlinear interaction of Fourier modes of different
wavelengths.  Unlike the crests of sinusoidal waves, we observe small
ripples at the wave peaks of the periodic wave at $s_1$.  As the
amplitude of the periodic solution increases, this nonlinear feature
is more pronounced.  For the periodic solution at $s_2$, near $x =
2\pi n$ for $n\in \mbb Z$, there is a flat plateau with wave peaks
shifted to the edges of the plateau, forming an interesting ``cat
ears'' structure.  These nonlinear features at the wave crests can be
attributed to the effect of the capillary force.  In panel~(b) of
Figures \ref{fig:wilton_small} and \ref{fig:wilton_large}, we show
contour plots of torus functions of these periodic traveling waves.
We observe that the width of the yellow region is larger for the
higher-amplitude periodic wave; in correspondence, this wave possesses
wider wave crests.

We compute secondary quasi-periodic bifurcation branches that
intersect the primary periodic branch at $s_1$ and $s_2$ and show the
corresponding results in Figures \ref{fig:wilton_small} and
\ref{fig:wilton_large}, respectively.  In both computations, we set
$M_x = 300, M_y = 150, N_x = 100, N_y = 50$ and use $\hat \eta_{0,1}$
as the continuation parameter.  We follow the two quasi-periodic
branches until $\hat{\eta}_{0,1} = 7\times 10^{-5}$ and
$\hat{\eta}_{0,1} = 1.1\times 10^{-4}$, respectively; the
corresponding quasi-periodic traveling waves are plotted with red
lines in panel (a) of Figures \ref{fig:wilton_small} and
\ref{fig:wilton_large}. The objective function is minimized to
$2.14\times 10^{-27}$ and $5.03 \times 10^{-28}$, respectively, for
these solutions.  In panel~(a) of Figure \ref{fig:wilton_small}, the
oscillations at the troughs of the quasi-periodic wave are ahead of
the ones of the periodic wave near $\xi = 3\pi, 5\pi, 7\pi, 11 \pi$
and are behind near $\xi = \pi$, which demonstrates the quasi-periodic
feature of the secondary bifurcation.

We also observe that the amplitude of the quasi-periodic wave in panel
(a) of Figure \ref{fig:wilton_small} is noticeably larger than the
periodic wave due to the activation of Fourier modes in the
quasi-periodic direction.  In panel~(c) of Figures
\ref{fig:wilton_small} and \ref{fig:wilton_large}, we show contour
plots of the torus functions of the quasi-periodic traveling waves in
panel (a). Unlike the periodic solution, the quasi-periodic solution
depends on $\alpha_2$. For example, one can see the variation of the
yellow and blue regions in the $\alpha_2$ direction.  Moreover, this
variation is rather oscillatory in Figure~\ref{fig:wilton_small},
which adds to the difficulty of computing higher-amplitude
quasi-periodic waves on the bifurcation branch.  The 1D quasi-periodic
waves are obtained by evaluating the corresponding torus functions
along the the red dashed line of slope $1/\sqrt{2}$.  In panel~(a) of
Figures~\ref{fig:wilton_small} and \ref{fig:wilton_large}, there will
be crests if the dashed line in panel (c) passes through the yellow
region and troughs if it passes through the blue region.  Due to the
variation in yellow region, the widths of the crests of the
quasi-periodic wave are no longer constant. For example, in panel (a)
of Figure~\ref{fig:wilton_large}, the crests of the quasi-periodic
wave are wider than those of the periodic wave near $\xi = 6\pi, 8\pi$
and narrower near $\xi = 4\pi, 10\pi$.


\section{Conclusion}

In this paper, we have presented a numerical study of two-dimensional
finite-depth free surface waves in the spatially quasi-periodic
setting. Specifically, we have studied both the initial value and
traveling wave problems.  For the initial value problem, we derived
the governing equations of water waves in the presence of a background
flow and a non-flat bottom boundary in conformal space.  As noted in
Remark \ref{rmk:periodic}, the
derivation is valid in both the quasi-periodic and periodic settings.
Motivated by the experiments of Torres \emph{et al.}
\cite{torres2003quasiperiodic} studying spatially quasi-periodic
surface waves in the presence of a quasi-periodic bottom boundary, we
computed the time evolution of an initially flat surface with a
background flow over a quasi-periodic bottom boundary. We also find
that the waves develop quasi-periodic patterns in which the distance
between adjacent wave peaks is not constant.

Next we computed spatially quasi-periodic traveling waves that
bifurcate from the zero-amplitude wave or from finite-amplitude
periodic traveling waves. Motivated by observations in
\cite{quasi:trav, quasi:bifur} that the Fourier coefficients of
quasi-periodic traveling waves decay slower along certain directions,
we derived the weakly nonlinear equations governing small-amplitude
quasi-periodic traveling waves in Section~\ref{sec:weak:approx} and
found that there is a divisor $\hat{S}_{j_1,j_2}$ in the formula
for the Fourier coefficients $\hat{\eta}_{j_1,j_2}$ of the weakly
nonlinear solutions. For example, in the case of deep water, this
divisor reads
\begin{equation}
\hat{S}_{j_1, j_2} = \frac{g}{k}(|j_1+kj_2| - k)(|j_1 + k j_2|-1).
\end{equation}
Due to the unboundedness of $1/\hat{S}_{j_1, j_2}$, the Fourier
coefficients along directions $|j_1+kj_2| - k = 0$ and $|j_1 + k
j_2|-1 = 0$ are expected to decay slower than in other directions.
We also study these divisors in the case of shallow water and
find that weakly nonlinear theory breaks down faster when $h$ is
smaller due to the factor of $h^4$ in the formula \eqref{eq:S:asym:h}
for $\hat S_{j_1,j_2}$.

In the current work, we assume that the bottom boundary remains fixed
in time.  In the future, we plan to further extend our method to study
quasi-periodic flows with a free surface over a moving bottom
boundary. In the case of periodic water waves, this has been studied
in \cite{ruban2005water, ruban2020waves}.  We also plan to analyze the
linear stability of periodic traveling waves
\cite{nicholls2011spectral, deconinck2011instability, tiron2012linear,
  pierce1994modulational} and investigate the long-time dynamics of
traveling waves under unstable subharmonic perturbations.  In the
quasi-periodic setting, we are able to compute the exact time
evolution of these perturbed waves instead of their linearized
approximations \cite{janssen2003nonlinear}.  We are also interested in
developing numerical methods, such as the Transformed Field Expansion
method \cite{nicholls:reitich:01,nicholls:reitich:06,qadeer}, to study
the dynamics of these waves in three dimensions where the conformal
mapping method no longer applies.  On the theoretical side, a rigorous
proof of the existence of quasi-periodic traveling waves is still an
open problem.  We expect it will be necessary to employ a Nash-Moser
iteration to tackle the small divisor problem, which has been
successfully used to prove the existence of temporally quasi-periodic
standing waves and traveling waves
\cite{berti2016quasi,berti2021traveling}.

\vspace*{5pt}
{
\noindent
\textbf{Funding:} This work was supported in part by the National Science
  Foundation under award number DMS-1716560 and by the Department of
  Energy, Office of Science, Applied Scientific Computing Research,
  under award number DE-AC02-05CH11231; and by an NSERC (Canada) Discovery Grant.
}

\vspace*{5pt}
{
\noindent
\textbf{Declaration of interests:} The authors report no conflict of interest.
}




\begin{appendix}
\section{Computation of the conformal mapping from a infinite horizontal strip to the fluid domain} \label{sec:conformal:transform}
In practice, the initial condition of the water wave is usually given in physical space. 
Therefore, we need to compute the conformal mapping $z(w, t)$ to transform the initial condition from physical space to conformal space. As shown in (\ref{eq:y:form}) and (\ref{eq:x:form}), the conformal mapping is determined by $h$, $x_0$, $\td \eta^\sfc$ and $\td \eta^\btm$, where $x_0$ is fixed to be zero in the scope of this paper and $h$, $\td \eta^\sfc$, $\td \eta^\btm$ are obtained by solving the following equations,
\begin{equation}\label{eq:solve:conf:map}
\begin{gathered}
\mc R_1(\alpha_1, \alpha_2) = \td \eta^\sfc - \td \eta^{\sfc, \ph}(\alpha_1 + \td\xi^\sfc, \alpha_2 + k\td\xi^\sfc) = 0, \\ 
\mc R_2(\alpha_1, \alpha_2) = \td \eta^\btm - \td \eta^{\btm, \ph}(\alpha_1 + \td \xi^\btm, \alpha_2 + k\td\xi^\btm) = 0, \\ 
\td\xi^\sfc = H^{\coth}[\td\eta^\sfc] + H^{\csch}[\td\eta^\btm], \qquad \qquad 
\td\xi^\btm = - H^{\csch}[\td\eta^\sfc] - H^{\coth}[\td\eta^\btm],
\end{gathered}
\end{equation}
which come from (\ref{eq:eta:b:conf}) and (\ref{eq:xi:eta:hilb}).  Moreover, we enforce the constraint $h = \hat{\eta}^\sfc_{\bds 0} - \hat{\eta}^\btm_{\bds 0}$ discussed in Section \ref{sec:conf:map} and rewrite $\td \eta^\btm$ as
\begin{equation}
\td \eta^\btm = \hat \eta^\sfc_{\bds 0} - h + P[\td \eta^\btm].
\end{equation}
Otherwise problem (\ref{eq:solve:conf:map}) is underdetermined and the solution is not unique.

In our computations, we consider $h$ and the Fourier coefficients of $\td \eta^\sfc$ and  $\td \eta^\btm$ as unknowns and define the following objective function
\begin{align}\label{eq:object:conf:map}
\mc F[h, \hat \eta^\sfc, \hat \eta^\btm] :&= \frac{1}{8\pi^2}\int_{\mathbb{T}^2}
  \mathcal{R}_1^2[h, \hat \eta^\sfc, \hat \eta^\btm] + \mathcal{R}_2^2[h, \hat \eta^\sfc, \hat \eta^\btm]
  \, \, d\alpha_1\,d\alpha_2 \\ 
 &\approx \frac{1}{2M_1M_2}\sum_{m_2 = 0}^{M_2-1} \sum_{m_1 = 0}^{M_1-1}
 \Big[\mc R^2_1(2\pi m_1/M_1, 2\pi m_2/M_2) + \mc R^2_1(2\pi m_1/M_1, 2\pi m_2/M_2)\Big].
 \notag
\end{align}
We apply a Levenberg-Marquardt method \cite{wilkening2012overdetermined} to solve the nonlinear least-squares problem~(\ref{eq:object:conf:map}) and compute the derivative of $\mc R_1$ and $\mc R_2$ with respect to the unknowns using the following variational equations
\begin{equation}\label{eq:phys2quasi}
\begin{gathered}
\delta \mc R_1 = \delta \td \eta^\sfc - \td\eta^{\sfc, \ph}_{x}\delta \td \xi^\sfc, \qquad 
\delta \mc R_2 = \delta  \hat\eta^\sfc_{\bds 0} - \delta h + P[\delta \td \eta^\btm]
- \td\eta^{\btm,\ph}_{x}\delta \td \xi^\btm,\\ 
\delta \td \xi^\sfc= H^{\coth} [\delta \td \eta^\sfc] + H^{\csch}[\delta \td \eta^\btm] +
\big(\delta H^{\coth}\big) [\td \eta^\sfc] + \big(\delta H^{\csch}\big)[\td \eta^\btm], \\ 
\delta \td \xi^\btm = -H^{\csch} [\delta \td \eta^\sfc] - H^{\coth}[\delta \td \eta^\btm] -
\big(\delta H^{\csch} \big)[\td \eta^\sfc] - \big(\delta H^{\coth}\big)[\td \eta^\btm].
\end{gathered}
\end{equation}
Here $\partial_x = \partial_{x_1} + k\partial_{x_2}$ and the symbols of $\delta H^{\coth}$ and $\delta H^{\csch}$
are 
\begin{equation}
\delta \hat{H}^{\coth}_{j_1, j_2} = \frac{i(j_1+kj_2)\delta h}{\sinh^2((j_1+kj_2)h)}, \qquad
\delta \hat{H}^{\csch}_{j_1, j_2} = -i(j_1+kj_2) \coth((j_1+kj_2)h)\csch((j_1+kj_2)h) \delta h.
\end{equation}

\end{appendix}




\bibliographystyle{abbrv}

\bibliography{refs}

\end{document}